Olival Freire Jr.∗

**Science and exile: David Bohm, the hot times of the Cold War, and his struggle for a new interpretation of quantum mechanics[#]**

ABSTRACT: In the early 1950s the American physicist David Bohm (1917-1992) lived through a critical moment, when he produced a new interpretation for quantum mechanics and had to flee from the United States as McCarthyism's victim. At that time he faced vicissitudes related to the Cold War and his move from Princeton to São Paulo. This article focuses on the reception among physicists of his papers on the causal interpretation, which were poorly received at the time. I describe his Brazilian exile and analyze the culture of physics surrounding the foundations of quantum mechanics. This article takes into account the strength of the Copenhagen interpretation of quantum physics among physicists, the way in which issues concerning the foundations of quantum mechanics were present in the training of physicists, the low status of these issues on research agendas, and the kind of results Bohm and collaborators were able to achieve. It also compares the reception of Bohm's ideas with that of Hugh Everett's interpretation. This article argues that the obstacles growing from the cultural context of physics at that time had a more significant influence in the reception of Bohm's ideas than did the vicissitudes related to the McCarthyist climate.

WHEN DAVID BOHM and Richard Feynman concluded their physics dissertations, during the first half of the 1940s, they were among the most promising students of their generation in American physics. After the war, however, their fates were very different. Feynman reached the zenith of his scientific career in the United States. In 1951, Bohm was forced to leave the United States forever. Before that, he got a position in Princeton, carried out his research on plasma theory, and published a praised graduate textbook entitled *Quantum Theory*. At that moment, his life suffered two important changes. He became a victim of McCarthy's anticommunist hysteria and changed his research focus to a heterodox subject of physics, developing a causal interpretation of quantum mechanics opposing the standard position in that subject. Those two changes were a turning point in his personal and scientific life. In the McCarthy period, Bohm could not survive in American academia. The first scientific position he was able to get was in Brazil, but he never enjoyed his 3-year Brazilian exile. However, having had his passport confiscated by American officials, it was as a Brazilian citizen that Bohm left Brazil in January 1955 to take a position at the Technion in Haifa. Two years later he went to England, where he finally found a convenient place to pursue his research for the rest of his prolonged exile. His main scientific interests remained in the unorthodox subject he began to work on in 1951, but from 1970 on he shifted his approach to what he called "implicit order." [1]

∗ Dibner Institute, MIT-E-56-100, 38 Memorial Dr, Cambridge, MA 02139. On leave of absence from Federal University of Bahia, Brazil. (freirejr@ufba.br). A first version of this paper was presented at the workshop, "Migrant Scientists in the Twentieth Century", Università degli Studi de Milano, Milan, June 20-22, 2003. I would like to thank the Dibner Institute for the History of Science and Technology, CNPq (Grant 303967/2002-1), CAPES, FAPESB, American Institute of Physics, and American Philosophical Society, for grants that have permitted me to finish this work. The staffs of the listed archives were instrumental for my archival research and for authorizing quotations. I am indebted to a number of people who read or commented on previous versions of this work, Michel Paty, Sam Schweber, Joan Lisa Bromberg, David Kaiser, Shawn Mullet, Alexei Kojevnikov, Alexis De Greiff, Paul Forman, Amy Ninetto, Leonardo Gariboldi, Osvaldo Pessoa, Amélia Hamburger, Saulo Carneiro, and Ileana Greca. Lilith Haynes, and the staff of the MIT Writing and Communication Center helped me with the English language. I use the following abbreviations: BP for Bohm Papers, Birkbeck College, University of London; AC for Arquivos do CNPq, Museu de Astronomia, Rio de Janeiro; RP for Rosenfeld Papers, Niels Bohr Archive, Copenhagen; WP for Wheeler Papers, American Philosophical Society, Philadelphia, USP for University of São Paulo, and *PR* for *Physical Review*.
[#] Paper to be published in ***Historical Studies on the Physical and Biological Sciences [HSPS]***.

At the beginning of the 1950s, the context in which physics was being practiced was not favorable to research on the interpretation of quantum mechanics. Bohm's causal interpretation was poorly received among physicists; in some important cases it was even received with hostility, and just a few people supported his ideas. Both the vicissitudes of Bohm's life and the poor initial reception of his causal interpretation have attracted the attention of historians, philosophers, and physicists, and some of them have blamed the former for the latter. "The political atmosphere in the US at that time did not help rational debate and in consequence there was little discussion and the interpretation was generally ignored for reasons that had more to do with politics than science," stated the physicist Basil Hiley, who was Bohm's assistant. F. David Peat, a science writer and former Bohm's collaborator, also found the political explanation for Bohm's unfavorable reception appealing, but limited its force to the Princeton physics community. The historians Russel

---

[1] David Bohm (1917-1992) received his PhD in 1943, under Julius Robert Oppenheimer, at Berkeley, and Richard Feynman (1918 – 1988) completed his PhD in 1942, under John Archibald Wheeler, at Princeton. As evidence of their promise in American physics, I remark that both of them were among the two dozen elite physicists who participated in the 1947-1949 Shelter Island, Pocono, and Oldstone conferences, and that the very Wheeler who supervised Feynman's dissertation suggested that Princeton hire Bohm. For the conferences, Sam Schweber, "Shelter Island, Pocono, and Oldstone: The emergence of American quantum electrodynamics after World War II," *OSIRIS, 2* (1986), 265-302. For Bohm's hiring by Princeton, John Wheeler, *Geons, black holes & quantum foam: A life in physics* (New York, 1998), 215-6. David Bohm, *Quantum theory* (New York, 1951); ibid., David Bohm, "A suggested interpretation of the quantum theory in terms of 'hidden' variables," *PR, 85* (1952), 166-179 and 180-193.

Olwell and Shawn Mullet, in their interesting and comprehensive studies of Bohm's prosecution under McCarthyism and of Princeton's procedures towards Bohm, blamed Bohm's Brazilian exile for the poor reception of Bohm's causal interpretation.[2] The interest in such issues has grown probably because the second half of the 20[th] century witnessed a dramatic change in that context. From the early 1970s on, the subject matter of Bohm's works – the interpretation and foundations of quantum mechanics – became a field of intensive research; Bohm's first proposal, slightly modified to so-called Bohmian mechanics, enjoyed a larger audience in the 1990s; and, by the end of the century, he was considered one of the most gifted protagonists in the field of research whose creation he contributed. A sign of the prestige accorded Bohm and the field of the foundations of quantum mechanics can be found in the volume in honor of the centenary edition of *Physical Review* which includes commentaries on and reprints from the most important papers ever published in the leading journal of American physics. Chapter 14, edited by Sheldon Goldstein and Joel Lebowitz, is entitled "Quantum Mechanics", but all of its papers, including Bohm's paper on the causal interpretation, concern foundations of quantum mechanics. A photo of Bohm opens the chapter. Thus, the initial poor reception of Bohm's ideas requires historical explanation.[3]

My main point is that in order to put Bohm's physical ideas into its cultural context, one needs to include in that context not only his persona as a person prosecuted by McCarthyism and later on his Brazilian exile, but also the ways in which the subject matter of Bohm's work, the foundations of quantum mechanics, were being practiced at the time.

---

[2] Basil Hiley, "David Joseph Bohm," *Biogr. Mem. Fell. R. Soc. Lond.*, *43* (1997), 105-31, on 113; F. David Peat, *Infinite potential: The life and times of David Bohm* (Reading, 1997), 133; Russel Olwell, "Physical isolation and marginalization in physics - David Bohm's Cold War exile," *ISIS, 90* (1999), 738-56, on 750; and Shawn Mullet, "Political science: The red scare as the hidden variable in the Bohmian interpretation of quantum theory," Senior Thesis HIS679, University of Texas at Austin, unpublished paper (1999). Mullet, after contact with sources about Bohm's stay in Brazil, has changed his views; cf. Shawn Mullet, "Creativity and the mainstream: David Bohm's migration to Brazil and the hidden variables interpretation," unpublished paper, Workshop on "Migrant scientists in the twentieth century," (Milan, 2003).
[3] Olival Freire Jr., "The historical roots of 'foundations of physics' as field of research (1950-1970)," *Foundations of Physics, 34* (2004), 1741-1760. For the evolution of the number of citations of Bohm's original paper on causal interpretation, see the graphics [enclosed], based on Fabio Freitas and Olival Freire Jr., "Sobre o uso da web of science como fonte para a história da ciência," *Revista da SBHC, 1* (2003), 129-147. For a discussion of "Bohmian mechanics", James Cushing, Arthur Fine, and Sheldon Goldstein, eds., *Bohmian mechanics and quantum theory: An appraisal* (Dordrecht, 1996). Henry Stroke, ed., *The Physical Review: The first hundred years - a selection of seminal papers and commentaries* (New York, 1995).

By this I mean both the role played among physicists by the dominant view at the time, i.e. the interpretation of complementarity, as well as the ways in which physicists included or excluded, in their research agendas and teaching duties, issues related to the foundations and interpretation of quantum mechanics. In putting scientific ideas in their cultural contexts I neither juxtapose them nor find strong causal links between them; as remarked by Peter Galison and Andrew Warwick, "understanding science as a cultural activity […] means learning to identify and to interpret the complicated and particular collection of shared actions, values, signs, beliefs and practices by which groups of scientists make sense of their daily lives and work." These authors also noted, "this kind of approach has already been widely applied to the history of the experimental sciences, but the literature on the theoretical side is much less developed." This note affects the present study, since no connections between theories and experiments related to the foundations of quantum mechanics had been made in the early 1950s, and we need to understand the reception of a theoretical approach in a context lacking experimental links. The suggested approach of science as a cultural activity calls attention for the role of pedagogy in the production of science, an aspect I take into account in this paper.[4]

To present my case, I will proceed along two fronts. I will firstly focus on a segment of Bohm's vicissitudes, his Brazilian exile, which is not yet well analyzed in the available literature, to argue that Bohm found support in Brazil for his research program. This part is also a contribution to filling a gap pointed out by Alexis De Greiff and David Kaiser; according to them, "despite several recent studies, the problem of the construction of knowledge outside the leading centers of calculation and, consequently, of the globalization of knowledge, remains woefully understudied."[5] Bohm continued to work consistently on the causal interpretation, kept in contact with colleagues abroad, discussed his proposal with visitors from Europe and the United States, and profited from collaboration with Brazilian physicists in achieving some of the published results on the causal interpretation. Bohm's deeds in Brazil did not match the views he expressed in the letters he exchanged at the time, the latter reflecting his personality and the vicissitudes he faced rather than being a reliable picture of his Brazilian stay. Additionally, I will show that Bohm would have faced elsewhere many of the obstacles that he faced in Brazil while working on causal interpretation. The second part of the paper brings out more about his activities in the 1950s, moving from the examination of Bohm's Brazilian exile to other dimensions related to the reception of his interpretation. The focus of the analysis is on the dispute between supporters of causal and complementarity interpretations, and, in particular, on the discourse of the several participants in the dispute. Bohm's proposal represented the first alternative to the complementarity interpretation, and for this reason it was seen as a major challenge to the established interpretation. The balance of the dispute, in the 1950s, was favorable to supporters of complementarity, since they succeeded in characterizing the dispute as being of a philosophical nature, meaning that its subject was not a matter for professional physicists. I will show how the fact that Bohm and his collaborators did not get any new results both influenced, and was influenced by, that characterization. The label of philosophical controversy will be framed in a broader context related to the training of physicists in quantum mechanics and their research agenda. The fact that the various interpretations were absent from courses and textbooks on quantum mechanics, and the idea that the interpretation of quantum mechanics was a question already solved by the founding fathers of quantum mechanics reinforced the opinion, tacitly shared by most of the physicists, that foundational and interpretative questions were not worthy of being included in physics research agenda.

---

[4] Peter Galison and Andrew Warwick, "Introduction: Cultures of theory," *Studies in History and Philosophy of Modern Physics, 29B* (1998), 287-294. For the role of pedagogy, David Kaiser, "Cold War requisitions, scientific manpower, and the production of American physicists after World War II," *HSPS, 33(1)* (2002), 131-159; *Drawing theories apart: The dispersion of Feynman diagrams in postwar physics* (Chicago, 2005, forthcoming); David Kaiser, ed., *Pedagogy and the practice of science: Historical and contemporary perspectives* (Cambridge, MA, 2005, forthcoming); and their bibliographical references.
[5] Alexis De Greiff and David Kaiser, "Foreword," *HSPS, 33(1)* (2002), 1-2.

David Bohm and his supporters challenged what Max Jammer called the "almost unchallenged monocracy of the Copenhagen school in the philosophy of quantum mechanics."[6] I will discuss the possible meanings and implications of Jammer's characterization, taking into consideration available studies on history and philosophy of quantum mechanics. Additionally, I will sketch two comparative essays: the first is related to the role of the McCarthyist climate in different contexts, while the second compares the reception of Bohm's causal interpretation with the interpretation suggested by Hugh Everett, in the second half of the 1950s.

---

[6] Max Jammer, *The philosophy of quantum mechanics: The interpretations of quantum mechanics in historical perspective* (New York, 1974), 250.

Hence, I conclude that when one looks for a balanced view of the poor reception enjoyed by the causal interpretation in the 1950s, factors other than the vicissitudes faced by Bohm emerge as more influential. These concern the way in which complementarity supporters reacted to Bohm's challenge, physicists practiced the foundations of quantum physics, and Bohm and his collaborators achieved research results. He could not find much support in that context, and for the same reason, I suggest that the later resonance of Bohm's ideas among physicists, and even his persona in the 1990s, only can be understood in terms of the changed cultural context from the 1970s on. In addition, I will show that by this ruse of history, Bohm's hidden variable interpretation, so poorly received in the 1950s, produced a byproduct which has been the main scientific achievement behind the changed landscape of the research on the foundations of quantum mechanics: the Bell theorem, published in 1965 by the Irish physicist John S. Bell.

1. BRAZILIAN EXILE

**The vicissitudes of a life during the Cold War**

The beginning of the 1950s was a turning point in Bohm's personal and scientific life, when he was diverted from the feverish path of a promising career in post-war American physics. The House Un-American Activities Committee [HUAC] subpoenaed him to talk about his activities and links with the Communist Party during the war at the Radiation Laboratory at Berkeley. He decided not to answer on these activities, claiming the Fifth Amendment, which guarantees protection against self-incrimination. He was indicted by a jury for contempt of Congress, arrested, freed on bail, and eventually acquitted on May 31, 1951. In this acquittal, Bohm was favored by a decision of the Supreme Court reasserting the rights of the Fifth Amendment for those who were being called before congressional committees. However, his personal damages were irreversible. At the beginning of the trial, Princeton University had placed him on paid leave, but when the moment to renew his appointment came, in June 1951, Princeton did not reappoint him. Princeton's decision still stirs up controversy and passion. John Archibald Wheeler wrote in his autobiography, after reminding the reader that it was he who had invited Bohm to Princeton, that "since the Bohm affair – which understandably polarized the campus – occurred while I was away, I played no part in it. Had I been there, I'm not sure I would have been outspoken in Bohm's defense. […] The university was gauche in its manner of dealing with Bohm, yet I could sympathize with its goal, to preserve its reputation as a center of unbiased scholarly inquiry, not the home of blind loyalty to one ideology or another."[7]

Feeling that he could not get another academic position in McCarthy's America, Bohm began to look for a position abroad. Especially in his attempt to get a position in Manchester,[8] he was supported by Albert Einstein, to whom he had become very close while discussing the interpretation of quantum mechanics. When Manchester did not hire him Brazil entered Bohm's scientific life, and not completely by chance. By that time, Princeton had graduated a small group of Brazilian physicists, and had become a meeting place for Brazilian physicists in the US [See photo, enclosed]. The Brazilians were Jayme Tiomno, who had graduated under John Wheeler & Eugene Wigner in 1950; José Leite Lopes, who had studied under Wolfgang Pauli & Josef Jauch in 1946 and been named a Guggenheim Fellow in 1949; and Walter Schützer, who had completed a Master degree in 1949. Bohm was one of the readers of Tiomno's doctoral dissertation and served as the chairman of his dissertation committee when Wigner was away.[9] Thus Bohm was already close to Brazilian physicists in Princeton. He asked Tiomno about job possibilities in Brazil, and Tiomno invited him to the University of São Paulo. The project of Bohm's going to Brazil engaged several people, with recommendation letters from Einstein and Oppenheimer, among others. The support of Abrahão de Moraes, then the head of the Physics Department, and Aroldo de Azevedo, a geographer who referred Bohm's application to the university board, were instrumental.[10] Later, to keep Bohm safely in his Brazilian position, de Moraes asked Einstein to send letters for use with the Brazilian authorities, including President Getúlio Vargas.[11]

---

[7] Those events are described in Olwell, Mullet [1999], Peat, 90-103, and Hiley, 113-114; all of them (ref. 2). Wheeler (ref. 1), 215. The most accurate description of Bohm's links with the Communist Party of the United States is Alexei Kojevnikov, "David Bohm and collective movements," *HSPS, 33(1)* (2002), 161-92. Oppenheimer's behavior in Bohm's case is discussed in David Cassidy, *J. Robert Oppenheimer and the American century* (New York, 2005), 282. For McCarthyism and universities, Ellen Schrecker, *No ivory tower: McCarthyism & the universities* (New York, 1986); and Jessica Wang, *American science in an age of anxiety: Scientists, anticommunism & the cold war* (Chapel Hill, 1999).
[8] Albert Einstein to Patrick Blackett, 17 Apr 1951, Albert Einstein Archives. I thank Michel Paty for sending me copies of Einstein's and Bohm's letters re Bohm's case.
[9] Jayme Tiomno, interviewed by the author, 4 Aug 2003.

### Brazilian exile and citizenship

Although it welcomed him, Brazil was never Bohm's first choice; it was the only possibility at a moment when he had no academic position and was scared of the growing McCarthyism in the United States. Bohm's Brazilian exile would be a period of marked uneasiness, as evidenced in his letters to Einstein, Hanna Loewy, Miriam Yevick, and Melba Phillips. He wrote, "I am afraid that Brazil and I can never agree,"[12] and reading those letters, one might conclude that Bohm's troubles in Brazil were mainly due to characteristics of the country – noise, dirt, heat – and what seemed to him a dullness in the intellectual atmosphere.[13] Bohm went to Brazil an innocent and, as soon as he arrived, he wrote optimistically to Einstein, "The university is rather disorganized, but this will cause no trouble in the study of theoretical physics. There are several good students here, with whom it will be good to work." Later, however, he expressed both his surprise and lack of prior knowledge about the country: "*The country here is very poor and not as advanced technically as the U.S., nor is it as clean.*"[14]

Bohm arrived in Brazil on October 10, 1951. One month later he would suffer a new blow, when American officials confiscated his passport and told him that he could only retrieve it to return to his native country.[15] Bohm was worried about the meaning of that ruling, and he expressed it in the first letter he wrote to Einstein, "Now what alarms me about this is that I do not know what it means. The best possible interpretation is that they simply do not want me to leave Brazil, and the worst is that they are planning to carry me back because perhaps they are reopening this whole dirty business again. The uncertainty is certainly very disturbing, as it makes planning for the next few years very difficult."[16] Bohm's stay in Brazil, without a passport, changed his mood; he wrote to Phillips, a close friend: "Ever since I lost the passport, I have been depressed and uneasy, particularly since I was counting very much on this trip to Europe as an antidote to all the problems that I have mentioned."[17] Bohm's response to the confiscation of his passport was to seek Brazilian citizenship.[18]

---

[10] Record number 816/51 [microfilm], Archives of the Faculdade de Filosofia, Ciências e Letras, USP.

[11] Abrahão de Moraes did not need to use them. The letter to President Vargas is published in *Estudos Avançados*, [São Paulo] *21* (1994).

[12] David Bohm to Hanna Loewy, 6 Oct 1953, BP (C.39).

[13] As he put it in short: "Brazil is an extremely backward and primitive country." David Bohm to Albert Einstein, 3 Feb 1954. Albert Einstein Archives.

[14] David Bohm to Albert Einstein, Nov 1951, BP (C.10-11), emphasis added.

[15] One should remark that David Fox, a colleague of Bohm's at Berkeley, faced a similar constraint in Israel but refused to deliver his passport, and did not lose his citizenship. Stirling Colgate to George Owen (Deputy Director, Visa Office – US State Dept), 4 Nov 1966, BP (C.8).

The events behind Bohm's exile in Brazil reveal the network of international solidarity that was able to back him at a crucial moment of his life. One knot of the network was the intellectual brilliance and moral strength of the name of Albert Einstein. The other knot was the confluence of diverse factors forming the Brazilian scientific context. The role played by the young community of physicists was instrumental, but this support would not have succeeded without a political and institutional basis. Brazil had had, since 1934, a young but autonomous university, the USP. That such a university could exist was a testament to the power of the regional elite in the state of São Paulo, who sought cultural hegemony after the defeat of their state in the 1932 rebellion.[19] At the time that Bohm went to Brazil, the country was experiencing a new democracy which arose from the anti-dictatorial struggle of the Brazilian people and also from the participation of Brazil in World War II on the Allied side, which had involved sending troops to the battles in Italy.[20] Only by considering that scientific, political and institutional context can one grasp how it was possible for the most important Brazilian university to welcome an American scientist tainted as a Communist by McCarthyism. Bohm's double identity as Marxist and Jew was not unfavorable in Brazil; on the contrary, this condition probably enlarged the desire of people to help him. The fragile Brazilian democracy initially legalized the Communist Party, but when the party was later banished from public life, Communists continued to play a role in Brazilian life. Indeed, among the Communist intellectuals one cand find the names of Jorge Amado and Graciliano Ramos, writers; Cândido Portinari, painter; Caio Prado Jr., historian; Mário Schönberg, physicist; and Oscar Niemeyer, architect.[21]

[16] David Bohm to Albert Einstein, Dec 1951, BP (C.10-11).
[17] David Bohm to Melba Phillips, *n.d.*, BP (C.46 - C.48).
[18] After Bohm became a Brazilian citizen, the US State Department decreed that Bohm had lost his American citizenship. He only recovered it in 1986, after a lawsuit.
[19] "It [creation of the USP] meant a political choice of São Paulo, after its defeat in the Constitutionalist Revolution of 1932, betting on science and culture as sources of its political redemption." Shozo Motoyama, "Os principais marcos históricos da ciência e tecnologia no Brasil," *Revista da Sociedade Brasileira de História da Ciência, 1* (1985), 41-49, on 44.
[20] Thomas Skidmore, *Politics in Brazil, 1930-1964; an experiment in democracy* (New York, 1967); and Antonio M. de Almeida Jr., "Do declínio do Estado Novo ao suicídio de Getúlio Vargas," Boris Fausto, *História geral da civilização brasileira, tomo III, vol. 3: O Brasil republicano – sociedade e política (1930-1964)* (Rio de Janeiro, 1996), 225-255.

In South America, Brazil had been a "terre d'accueil" for Jews since the beginning of the 20th century, owing to the weakness of anti-Semitic feelings. The most important literary recognition of this tolerance may have come from the Austrian Jewish writer Stefan Zweig (1881-1942), who argued, in 1941, that Brazil could be a land of future to the Jews persecuted in Europe. A sociological study about the Jewish community in São Paulo seems to confirm and explain that tolerance. Rattner identified the social and economic conditions associated with the exceptional phase of development in the metropolitan area of São Paulo as factors favorable to the integration of the Jewish community in that city, before and after World War II. This does not deny, however, the existence of anti-Semitic features in the nationalist politics of the dictatorial regime of the "Estado Novo" (1930-1945); and the weakness of anti-Semitic feelings in Brazil should not lead one to endorse the myth of Brazilian racial democracy as far as its African descendants are concerned.[22]

So, it is not by chance that one of the most gifted Brazilian physicists – Mario Schönberg – a member of the same Physics Department into which Bohm was accepted, was a Jew who had been a representative of the Communist party in the state parliament after World War II. Olwell remarks that "the political situation in Brazil was by no means free of fear either," and cites Schönberg's arrest in 1948. However, to understand how Bohm could teach at USP while he could not teach at Princeton, we need to look closely at the peculiarities of the fragile Brazilian democracy at that time. There were arrests, even illegal ones, of political activists. Unions and the Communist Party were closed down.

---

[21] Leôncio M. Rodrigues, "O PCB: os dirigentes e a organização," Boris Fausto (ref. 26), 361-443, on 412.
[22] Stefan Zweig, *Brazil: a land of the future* (Riverside, 2000), [1st ed. 1941]; Henrique Rattner, *Tradição e mudança (a comunidade judaica em São Paulo)* (São Paulo, 1977); Maria L. T. Carneiro, *O anti-semitismo na era Vargas (1930-1945)* (São Paulo; Thomas Skidmore, *Black into white; race and nationality in Brazilian thought* (Durham, 1993). I am grateful to Marcos Chor and Augusto Videira for some remarks about the literature on this subject.

However, Communist intellectuals, like the physicist Schönberg and others, could keep their positions at the universities. This could not have happened 20 years later, during the military dictatorship [1964-1985], when Schönberg was arrested and forced to retire by presidential decree. Obliged retirement was also the fate of liberals such as the physicists Tiomno, Leite Lopes, the sociologist Fernando Henrique Cardoso, who would later be President of Brazil, and the economist Celso Furtado.[23]

### Bohm's interpretation of quantum mechanics

I need now to present Bohm's hidden variable proposal to get a better idea about the issues at stake, and as his proposal was closely connected to the models he built, we need to have a rough idea of them. He depicted quantum systems, such as electrons, as particles with well defined positions, and associated them with a general function to be written in the form $\psi = R \exp (iS/\hbar)$. Substituting such a function in the Schrödinger equation and exploiting analogies between the equations resulting of this substitution and the Hamilton-Jacobi equations of classical mechanics, he derived that such electrons should have a well defined momentum given by the expression $\mathbf{p} = \nabla S(\mathbf{x})$. In addition, the electrons were submitted to a "quantum potential" $U(\mathbf{x}) = -\hbar^2\nabla^2R/2mR$, in addition to the potentials known from classical physics. In this model, $P = |\psi(\mathbf{x})|^2$ gives the probability density of a statistical ensemble of particle positions. At this point, one should note that electrons in Bohm's models have well defined positions and momenta; thus, they have continuous and well defined trajectories. As the complementarity interpretation of quantum mechanics prevents the simultaneous definition of such variables, their joint existences are the hidden variables in Bohm's models. They are "hidden" in the usual interpretation of quantum mechanics. Bohm went ahead and developed his approach in a twofold movement. He considered that the measurement devices also included well defined positions and momenta, and obtained the observable results writing and analyzing the Hamiltonian of the coupling between such devices and the systems. In the second movement, he applied those ideas to detailed calculations of a number of simple quantum systems such as stationary states, transitions between stationary states, including scattering problems, and the Einstein-Podolsky-Rosen *gedankenexperiment*. To achieve results compatible with those from quantum mechanics, he considered that a photon in its interaction with matter was not a particles, but a bunch of electromagnetic waves.

---

[23] Olwell (ref. 4), 750; Rodrigues (ref.27); for the last period, see Thomas Skidmore, *The politics of military rule in Brazil: 1964-85* (New York, 1988), especially chapter 4.

The results Bohm obtained were impressive. For phenomena in which relativistic considerations were not necessary, his treatment led to a full equivalence with usual quantum mechanical predictions, but using models based on assumptions in flagrant conflict with the complementarity interpretation of quantum mechanics. Bohm's interpretation departed from complementarity, or the "usual interpretation" as he called it, in its essential assumption, i.e. "that the most complete possible specification of an individual system is in terms of a wave function that determines only probable results of actual measurement processes." In addition, he promised that some assumptions of his models could be relaxed and could permit successful predictions different from quantum mechanics in domains in which this theory was facing difficulties or new phenomena, such as the myriad of new particles recently discovered as well as infinities in quantum electrodynamics. According to Bohm, "the usual mathematical formulation seems to lead to insoluble difficulties when it is extrapolated into the domain of distances of the order of $10^{-13}$ cm or less. It is therefore entirely possible that the interpretation suggested here may be needed for the resolution of these difficulties." Finally, we should note that Bohm was aware that his quantum potential exhibited strange features, such as the instantaneous propagation of interactions in systems with many bodies. However, he hoped that feature would disappear in a future relativistic generalization of his models; this hope was also a commitment to find such a generalization.[24]

Even before Bohm's papers appeared in print, Albert Einstein and Wolfgang Pauli remarked to him that Louis de Broglie had suggested a similar approach in 1927, a fact of which Bohm was not aware. Pauli's remarks were more critical, for a number of reasons. He had criticized de Broglie's approach at that time, and de Broglie had given up his own proposal; now it was up to Bohm to face the same criticism. In addition, Pauli's remarks stirred up a dispute regarding priorities. Pauli's criticism was that de Broglie's proposal fitted Max Born's probabilistic interpretation of the $\psi$ function only for elastic collisions, and illustrated his argument with the scattering of particles by a rotator, a problem which had been solved by Enrico Fermi a year before. Pauli showed that de Broglie's proposal did not provide the rotator stationary states before and after the scattering, and had considered this failure something intrinsic to de Broglie's assumption of particles with determined trajectories in the usual space-time. Pauli's updated criticism challenged Bohm to exhibit the strengths of his approach.[25]

---

[24] Bohm (ref. 1).

Pauli's criticisms were addressed to the first version of Bohm's paper, which Bohm had sent to him asking for critical comments. Bohm took into account Pauli's remarks and gave up some features of his models and worked out to their full consequences his first ideas, and these changes made the difference in comparison with de Broglie's earlier works. The first version of Bohm's paper did not survive, but one can have a glance at it in Bohm's letter responding to Pauli's criticisms: "I hope that this new copy of will answer some of the objections to my previous manuscript. […] To sum up my answer to your criticisms […], I believe that they were based on the excessively abstract assumptions of a plane wave of infinite extent for the electrons $\Psi$ function. As I point out in section 7 of paper I, if you had chosen an incident wave *packet* instead, then after the collision is over, the electron ends up in one of the outgoing wave packets, so that a stationary state is once more obtained." Pauli did not read the second manuscript as he considered it too long, which Bohm did not like: "If I write a paper so 'short' that you will read it, then I cannot answer all of your objections. If I answer all of your objections, then the paper will be too 'long' for you to read." Bohm warned Pauli – "I really think that it is your duty to read these papers carefully" – but as a precaution wrote a long letter to Pauli explaining his views and the improvements he had made to his models. One of these letters summarizes these improvements:[26]

---

[25] Einstein's remark is in Michel Paty, "Sur les 'variables caches de la mécanique quantique – Albert Einstein, David Bohm et Louis de Broglie," *La Pensée*, 292 (1993), 93-116. Bohm to Pauli, [Jul 1951], Wolfgang Pauli, *Scientific Correspondence*, Vol IV – Part I, [ed. by Karl von Meyenn], (Berlin, 1996), 343-5. Most of Pauli's letters to Bohm did not survive; we infer their contents from Bohm's replies. Bohm to Karl von Meyenn, 02 Dec 1983, ibid, on 345. Broglie's pilot wave and Pauli's criticisms are in *Électrons et Photons – rapports et discussions du cinquième conseil de physique* (Paris, 1928), 105-141, and 280-2.
[26] Bohm to Pauli, [Jul 1951], [Summer 1951], [Oct 1951], [20 Nov 1951]; Pauli, (ref. 25), on 343-6, 389-94, and 429-462.

In the second version of the paper, these objections are *all* answered in detail. The second version differs considerably from the first version. In particular, in the second version, I do not need to use "molecular chaos." You refer to this interpretation as de Broglie's. It is true that he suggested it first, but he gave it up because he came to the erroneous conclusion that it does not work. The essential new point that I have added is to show in detail (especially by working out the theory of measurement in paper II) that his interpretation leads to all of the results of the usual interpretation. Section 7 of paper I is also new [transitions between stationary states – the Franck-Hertz experiment], and gives a similar treatment to the more restricted problem of the interaction of two particles, showing that after the interaction is over, the hydrogen atom is left in a definite "quantum state" while the outgoing scattered particle has a corresponding definite value for its energy.

Eventually, Pauli studied in detail Bohm's papers, "I just received your long letter of November 20 and I also studied more thoroughly the details of your paper." Together with this letter, the only from Pauli of which the manuscript survives, there was an appendix with some calculations on what Pauli called "de Broglie' 'streamline-quantum-force' image." Bohm won the dispute with Pauli, whom many considered the foremost critic among physicists in the 20[th] century. Pauli conceded that Bohm's model was logically consistent: "I do not see any longer the possibility of any logical contradiction as long as your results agree completely with those of the usual wave mechanics and as long as no means is given to measure the values of your hidden parameters both in the measuring apparatus and in the observed system." Pauli finished his statement with a caveat that would haunt Bohm and collaborators for many years, "as far as the whole matter stands now, your 'extra wave-mechanical predictions' are still a check, which cannot be cashed." Pauli would not end his opposition to the hidden variable interpretation, as we will see later, but for the moment, it is interesting to remark that in this very appendix was one of his later criticisms: "to ascribe $\Psi(x)$ 'physical reality' and not to $\varphi(p)$ destroys a transformation group of the theory." For Bohm, it was more difficult to settle the priority dispute with de Broglie, reiterated by Pauli, than to answer Pauli's physics criticism.[27]

Louis de Broglie had been working with the idea of double solution, that is, the waves resulting from the Schrödinger equation pilot the particles, but particles themselves are singularities of the waves. On the eve of the 1927 Solvay council he gave up this idea due to its mathematical difficulties, and presented his report to that meeting with just the "pilot wave," adding particles as objects external to the theory. Since the 1927 meeting he had given up his approach and adhered to the complementarity interpretation. Bohm was right in remarking that de Broglie had not carried his ideas to a logical conclusion, but it was also clear that de Broglie had a share, previously acquired, in the idea of the hidden variables in quantum mechanics. Bohm resisted accepting it. To Pauli he wrote this interesting analogy, just after summarizing to him what he had added to de Broglie's ideas: "If one man finds a diamond and then throws it away because he falsely concludes that it is a valueless stone, and if this stone is later found by another man who recognize its true value, would you not say that the stone belongs to the second man? I think the same applies to this interpretation of the quantum theory.[28]

---

[27] Pauli to Bohm, 03 Dec 1951, plus an appendix, Pauli, (ref. 25), 436-441.

Eventually, however, Bohm found a diplomatic way, which was suggested by Pauli, to recognize de Broglie's priority while maintaining the superiority of his own work: "I have changed the introduction of my paper so as to give due credit to de Broglie, and have stated that he gave up the theory too soon (as suggested in your letter)." In addition to changing the introduction he added an appendix to the second part of his paper for "a discussion of interpretations of the quantum theory proposed by de Broglie and Rosen," and also rebutted Pauli's criticisms. By the time Bohm's papers appeared in print, de Broglie was again shifting his own position, this time coming back to his 1926-1927 approaches, and together with his assistant Jean-Pierre Vigier, would become the most important of Bohm's allies in the hidden variable campaign.[29]

### Brazilian activities and the reception of the causal interpretation

In some recent accounts, Brazilian exile was, for Bohm, a period in which he abandoned research or was completely blocked from pursuing it. We will see later how far from reality these descriptions were, but for now let us cite just a few of these accounts. Jessica Wang wrote that McCarthyism forced him to give up research for several years. Later, she slightly modified her views: "Unhappy with the quality of intellectual life at the

---

[28] For the evolution of de Broglie's ideas, Louis de Broglie, *Nouvelles perspectives en microphysique* , (Paris, 1952), 115-143. Bohm to Pauli, [Oct 1951], Pauli (ref. 25).
[29] Bohm to Pauli [20 Nov 1951], Pauli (ref. 25). Bohm, (ref. 1). 191-3.

University of São Paolo [sic] and beset with physical ailments, Bohm searched for a way out." Olwell recognized that "Bohm continued to work on questions of theoretical physics," but he did not comment on the results of that work, and added "in isolation." Olwell considered that Israel, in contrast to Brazil, "was a supportive place for Bohm's work in physics," and quoted his work with Yakir Aharonov. Nevertheless, Olwell did not take into account that Bohm's main work with Aharonov (Aharonov–Bohm effect) did not concern the causal interpretation of quantum mechanics. It was rather the proposal of a new physical effect derived from quantum mechanics but still not recognized by physicists at the time. Olwell, taking correctly into account the level of experimental physics in Brazil, wrote that "the Brazilian physics community lacked the kind of tools Bohm had used as a graduate student in experimental physics." But he did not consider that, since before leaving the United States, Bohm had been completely dedicated to the problem of the foundations of quantum mechanics, a field of theoretical physics with no contact with experiments in the 1950s. Experiments in this field came out later.[30] In contrast to these views, and considering that almost all his scientific interests were focused on the causal interpretation, I will argue that Bohm developed an intense and large scientific activity in Brazil. He discussed his proposal with foreign visitors, like Richard Feynman, Isidor Rabi, Léon Rosenfeld, Mario Bunge, Carl Friedrich von Weizsäcker, Herbert Anderson, Donald Kerst, Marcos Moshinsky, A. Medina, and Guido Beck, and Brazilian physicists like Schönberg and Leite Lopes. And yet, most important of all, in Brazil his work led not only to some individual publications but also to papers in collaboration with foreign visitors, such as the Frenchman Jean-Pierre Vigier, who went to Brazil for three months especially to work with Bohm, the American Ralph Schiller, who had been a student of the cosmologist Peter Bergmann at Syracuse University and stayed in Brazil for two years as Bohm's assistant, and the Brazilians Tiomno and Walther Schützer.

Bohm's main hopes for getting an ally among foreign visitors in Brazil were directed towards Richard Feynman, who was spending his sabbatical year in 1951 at the CBPF in Rio de Janeiro.[31] Bohm liked the way Feynman initially reacted to his talk: "At the scientific conference at Belo Horizonte, I gave a talk on the quantum theory, which was well received. Feynman was convinced that it is a logical possibility, and that it may lead to something new." His interaction with Feynman reinforced his conviction that he needed to talk with physicists in order to convince them, and that in Brazil, without a passport, everything became more difficult. How large was Bohm's bet on Feynman can be inferred from this letter to Hanna Loewy, which is also evidence of Bohm's distrust of the current trends of physics at the time:

> Right now, I am in Rio giving a talk on the quantum theory. About the only person here who really understands is Feynman, and I am gradually winning him over. He already concedes that it is a logical possibility. Also, I am trying to get him out of his depressing trap down long and dreary calculations on a theory [procedures of renormalization in Quantum Field Theory] that is known to be of no use. Instead maybe he can be gotten interested in speculation about new ideas, as he used to do, before Bethe and the rest of the calculations got hold of him.

---

[30] Jessica Wang, "Science, security and the Cold War: The case of E.U. Condon," *ISIS, 83* (1992), 238-89, on 267; Wang (ref. 7), on 278; Olwell (ref. 2), on 750.
[31] José Leite Lopes, "Richard Feynman in Brazil: personal recollections," *Quipu, 7* (1990), 383-97; and Jagdish Mehra, *The beat of a different drum* (New York, 1994), 333-342.

Bohm's hopes were unfounded, since "in his physics Feynman always stayed close to experiments and showed little interest in theories that could not be tested experimentally." As we have discussed, at the time the hidden variable approach had no connections with experiments. Indeed, the only reference Feynman made to hidden variables approach was to include it as one of the possible avenues for the development of theoretical physics, in a general paper published in a Brazilian science journal. This minor reference was too little for Bohm's hopes.[32] Guido Beck, who was Heisenberg's assistant, and a refugee from the Nazi regime, was living in Brazil at the time Bohm stayed there. Bohm found Beck a supportive person for his scientific activities, even if Beck did not share a belief in the causal interpretation. Beck defended Bohm against the acrimonious criticisms of Rosenfeld – especially the comparison between Bohm and a tourist - and maintained that one should wait to see what physical results Bohm would be able to get. He was also instrumental in Bohm's relationship with the CNPq's scientific director, Costa Ribeiro, concerning funding for Bohm's research.[33] The Argentine Mario Bunge, who had

---

[32] David Bohm to Hanna Loewy, 1951, [the talk was in a meeting of the Sociedade Brasileira para o Progresso da Ciência]; David Bohm to Melba Phillips, 22 Oct 1951; David Bohm to Hanna Loewy, 4 Dec 1951, BP (C.38); Silvan S. Schweber, "Feynman, Richard," John Heilbron, ed., *The Oxford Guide to the History of Physics and Astronomy* (New York, 2005), 118-20; Richard Feynman, "The present situation in fundamental theoretical physics," *Anais da Academia Brasileira de Ciências*, 26 (1) (1954), 51-60. For the role played by Feynman, Bethe, and the renormalization calculations in physics at that time, see Sam Schweber, *QED and the men who made it: Dyson, Feynman, Schwinger, and Tomonaga* (Princeton, 1994).
[33] Rosenfeld was sensitive to Beck's remarks. In the English translation of the original French paper Rosenfeld deleted the comparison which had been criticized by Beck. The original expression is: "on

been a doctoral student of Guido Beck in Buenos Aires, is a lesser-known case of adhesion to the causal interpretation. He read Bohm's papers and became motivated to work in such a direction. Bohm replied to the letter in which he asked questions about the hidden variable models with an invitation to come to São Paulo. Bunge spent one year working with Bohm, but in spite of the good conversation, nothing came out. Indeed, Bunge attacked a problem which was more difficult than he had thought before, that is, the "Bohmization" of relativistic quantum mechanics and elimination of infinitudes in quantum electrodynamics. Besides Bunge, other causal interpretation supporters unsuccessfully tackled the same problem. In the middle of the 1960s, disenchanted with the hidden variable interpretation, he would give it up, as we will see later.[34] Bohm enjoyed conversation with Feynman, Beck, and Bunge, in addition to the cooperative work with Vigier and Schiller; however, his feelings were different with other visitors such as Isidor Rabi, Léon Rosenfeld, and von Weizsäcker.

> We had an international Congress of Physics. […] 8 physicists from the States (including Wigner, Rabi, Herb, Kerst, and others), 10 from Mexico, Argentina, and Bolivia, aside few from Europe, were brought here by the UNESCO and the Brazilian National Res. Council. […] The Americans are clearly very competent in their own fields, but very naïve and reactionary in other fields. […] I gave a talk on my hidden variables, but ran into much opposition, especially from Rabi. Most of it made no real sense.[35]

Bohm complemented his description of the meeting by formulating Rabi's view thus: "As yet, your theory is just based on hopes, so why bother us with it until it produces results. The hidden variables are at present analogous to the 'angels' which people introduced in the Middle Ages to explain things." We can be sure that Bohm produced a faithful description of the content of Rabi's intervention, even if the proceedings of the meeting do not include the reference to the medieval angels. Indeed, according to Rabi, "I comprend que le pionnier s'avançant dans un territoire inconnu ne trouve pas d'emblée la bonne route; on comprend moins qu'un touriste s'égare encore après que ce territoire a été levé et cartographié au vingt-millième." Léon Rosenfeld, "L'évidence de la complementarité," André George, ed., *Louis de Broglie - physicien et penseur* (Paris, 1953), 43-65, on 56; idem, "Strife about complementarity," *Science Progress,* 163 (1953), 393-410, reprinted in Robert Cohen and John Stachel, eds., *Selected papers of Léon Rosenfeld* (Dordrecht, 1979); Guido Beck to Léon Rosenfeld, 1 May 1952, RP. Rosenfeld to Beck, 9 Feb 1953; Bohm to Beck, 16 Sep 1952; 31 Dec 1952; 13 Apr 1953; 5 May 1953; 26 May 1953; Guido Beck Papers, Centro Brasileiro de Pesquisas Físicas, Rio de Janeiro. For a collective volume of historical and scientific papers in honor of Guido Beck, see *Anais da Academia Brasileira de Ciências*, 67, Supplement 1 (1995).
[34] Mario Bunge to the author, 01 Nov 1996, and 12 Feb 1997.
[35] David Bohm to Miriam Yevick, [received 20 Aug 1952]; Bohm to Melba Phillips, *n.d.*, BP. I merged the two letters in my narrative. I am grateful to Shawn Mullet for the courtesy of a CD with copies of Miriam Yevick's letters. They are now also available at BP.

do not see how the causal interpretation gives us any line to work on other than the use of the concepts of quantum theory. Every time a concept of quantum theory comes along, you can say yes, it would do the same thing as this in the causal interpretation. But I would like to see a situation where the thing turns around, when you predict something and we say, yes, the quantum theory can do it too." Bohm's main answer was to compare the current context with the debates on atomism in the 19[th] century: "[E]xactly the same criticism that you are making was made against the atomic theory – that nobody had seen the atoms, nobody knew what they were like, and the deduction about them was gotten from the perfect gas law, which was already known." The "much opposition" Bohm referred to included questions related to the relativistic generalization of his model and its experimental predictions. Indeed, Anderson, from Chicago, wanted to know how Bohm could recover the quantum feature of indiscernibility of particles, i.e., the exclusion principle; the Mexican physicist Medina asked if Bohm's approach could "predict the existence of a spin of a particle as in field theory;" the Brazilian Leite Lopes and the American Kerst called for experiments which could contrast both interpretations; and Moshinsky, from the University of Mexico, posed the question whether there is "reaction of the motion of the particle on the wave field." Bohm's answer to Anderson is interesting in that he emphasized that the causal interpretation only needed to reproduce the experimental predictions of quantum theory, but not each one of its concepts, "All I wish to do is to obtain the same experimental results from this theory as are obtained from the usual theories, that is, it is not necessary for me to reproduce every statement of the usual interpretation. […] You may take the exclusion principle as a principle to explain these experiments [levels of energy]. But another principle would also explain them.[36]

Rosenfeld was one of the main opponents of the causal interpretation, and for this reason, I will consider his case in the second part of this paper. For the moment, however, I remark only that conversation between Bohm and him on quantum mechanics would always be harsh. As he wrote to Bohm, "I certainly shall not enter into any controversy with you or anybody else on the subject of complementarity, for the simple reason that there is not the slightest controversial point about it." Rosenfeld went to Brazil especially motivated to discuss the epistemological problems of quantum mechanics. He gave a course on classical statistical mechanics in Rio de Janeiro, published papers in Portuguese on the epistemological lessons of quantum mechanics, and gave a talk in São Paulo on the non-controversial issue of complementarity. Bohm was not completely disappointed with the debate, at least as he reported to Aage Bohr: "Prof. Rosenfeld visited Brazil recently, and we had a rather hot and extended discussion in São Paulo, following a seminar that he gave on the foundations of the quantum theory. However, I think that we both learned something from the seminar. Rosenfeld admitted to me afterwards that he could at least see that my point of view was a possible one, although he personally did not like it."[37] The discussions between von Weizsäcker and Bohm left contradictory records. The German physicist recognized in 1971 that debates with Bohm on hidden variables were at the origin of his motivation to work on what he called "complementarity logic," which is a case of many-valued logic. Bohm, however, was strongly influenced by the fact that von Weizsäcker allied himself in São Paulo with a group of physicists with whom Bohm was in dispute at the Physics Department, a dispute rather more related to positions and funding than to Bohm's causal interpretation. He saw von Weizsäcker's activities as a plot, and asked support from his friends, the physicist Philip Morrison and the mathematician Miriam Yevick, warning that "Nazis [are] taking over Brazilian physics," and suggesting, "try to see what you can do about lining up publicity against Weissacre [sic], *but don't do a thing till I say 'go'*." To Guido Beck, he detailed who was in the group involved: "I am writing you to let you know that Marcello and Stammreich, *apparently* acting on behalf of the Weissacker – Leal [sic] group are doing their best to annoy me." Marcello is the Brazilian physicist Marcello Damy de Souza Santos. He worked with cosmic rays and had built in 1950 the USP's betatron, which was the first accelerator to be used in Latin America. The German spectroscopist Hans Stammreich had migrated to Brazil in the 1940s, and was a professor of physics at USP, and "Leal" refers to the brothers Jorge and Paulo Leal Ferreira, Brazilian physicists who eventually founded the Instituto de Física Teórica, in São Paulo. According to Bohm, Damy's arguments involved ideological considerations, since he had "been telling everyone here that (a) I am Communist, (b) My theory is Marxist," to which Bohm added, "both statements are, of course, nonsensical."

---

[36] Ibid. *New Research Techniques in Physics*, Proceedings, [Rio de Janeiro and São Paulo, July 15-29, 1952], Rio de Janeiro, 1954, pp.187-198.

As the dispute concerned the hiring of Bohm's assistants, he wrote that "Stammreich [had] accused [him] of filling the place with North Americans;" but added, "I was warmly defended by several Brazilians, however, and my proposal was passed by a large majority." Apparently, prosaic disputes took the place of more fruitful discussions between Bohm and von Weizsäcker.[38]

The physics produced by Bohm in Brazil was relevant to the development of the causal program, as one can infer from later accounts by the supporters of the program. In fact, Bohm's papers with Tiomno & Schiller, and his paper with Vigier, were always considered by Bohm as the main achievements of the causal program at that time.[39] With Vigier, Bohm met Wolfgang Pauli's early criticism that Bohm had included an arbitrary element in the causal interpretation, equaling its probability distribution to the function that satisfied the Schrödinger's equation of quantum mechanics.[40] Bohm had tried to solve the question by himself, without success,[41] and Louis de Broglie and Vigier had been sensitive to that criticism since the beginning of 1952.[42] In 1954, Bohm and Vigier were able to prove that, under some general conditions, any function could become a solution of the Schrödinger equation. To get that result, they used the analogy between Bohm's approach and the hydrodynamic approach to quantum mechanics, suggested by Erwin Madelung in 1926, and embedded the microscopic quantum particles in a subquantum medium with random fluctuations.[43] Thus, the "molecular chaos" that Bohm had abandoned after his discussions with Pauli came back in his work with Vigier. Since Tiomno was Bohm's main Brazilian collaborator, it is interesting to recover a recent quotation from Wheeler, which gives an idea of Tiomno's stature as a theoretical physicist: "I always think of Tiomno as one of the most unappreciated of physicists. His work on muon decay and capture in 1947-1949 was path-breaking and would still merit recognition by some suitable award." In 1987, Wheeler nominated Tiomno, Chien-Shiung Wu, Robert E. Marshak, and E. C. George Sudarshan for the Nobel Prize, without success.[44] With Tiomno and Schiller, Bohm included spin as a physical property of his model, although they used analogies with Pauli's equation and not relativity.[45] Still, with Vigier, Bohm began to approach elementary particles by using models of those particles as extended bodies in space-time, and equaling their freedom degrees to their quantum numbers, in order to get a classification for the myriad newly discovered particles. This paper was published some years later,[46] and it was the beginning of a lasting collaboration among Bohm, Vigier, de Broglie, and their associates.[47] With Walter Schützer, Bohm worked on a study of the role of probability in physical theories.[48] In Brazil, he also developed his philosophical ideas, changing the privileged status he had attributed to causal descriptions in physical theories. Those developments, as they appeared in *Causality and chance in modern physics* – a book written when still in Brazil but only published in 1957, led Bohm to conceive of both causal and probabilistic descriptions as possibilities with the same philosophical rank in science. This reflection contributed to take him far from his causal description, an intellectual shift that would clearly appear in Bohm's ideas of the 1970s.

---

[38] "In 1953, while still a member of the Max-Planck-Institute in Göttingen, von Weizsäcker visited, in an administrative capacity, Brazil where he met with David Bohm in São Paulo and discussed with him the problem of hidden variables. After his return to Göttingen von Weizsäcke, anxious to work out some ideas raised in his discussion with Bohm, decided to conduct a seminar, together with Georg Süssmann, with the objective of studying alternative formulations of quantum mechanics. It was in the course of this seminar, which was also attended by Heisenberg, that von Weizsäcker worked out his 'complementarity logic'." Jammer (ref. 6)376. Bohm to Melba Phillips (w/d), Bohm to Miriam Yevick (w/d), BP. Bohm to Guido Beck (w/d), Guido Beck Papers.

[39] David Bohm, *Causality and chance in modern physics* (London, 1984), 114, 118, notes 11 & 12; and David Bohm and Basil Hiley, *The undivided universe* (London, 1993), 205. Bohm's papers written in collaboration with other physicists while he was in Brazil were David Bohm and Jean-Pierre Vigier, "Model of the causal interpretation of quantum theory in terms of a fluid with irregular fluctuations," *PR, 96* (1954), 208-16; David Bohm, Ralph Schiller, and Jayme Tiomno, "A causal interpretation of the Pauli equation (A)." *Nuovo Cimento, Suppl.Vol 1* (1955), 48-66; David Bohm and Ralph Schiller, "A causal interpretation of the Pauli equation (B)." *Nuovo Cimento, Suppl. Vol 1* (1955), 67-91; David Bohm and Walter Schützer, "The general statistical problem in physics and the theory of probability," *Nuovo Cimento, Suppl Vol. 2* (1955), 1004-47. Besides, Bohm published five articles and letters alone. For an analysis of the ensemble of these papers, See Olival Freire Jr., *David Bohm e a controvérsia dos quanta* (Campinas, 1999).

[40] Wolfgang Pauli, "Remarques sur le problème des paramètres cachés dans la mécanique quantique et sur la théorie de l'onde pilote," André George, ed., *Louis de Broglie – physicien et penseur* (Paris, 1953), 33-42, 38.

[41] David Bohm, "Proof that Probability Density Approaches $|\psi|^2$ in Causal Interpretation of the Quantum Theory," *PR, 89* (1953), 458-66. A simplified and shortened version of this paper was presented at the above mentioned international scientific meeting held in Brazil, (ref. 36), 187-198.

[42] "C'était aussi un des problèmes décisifs que Bohm n'avait pas traité dans ses papiers de 1952. » Jean Pierre Vigier, interviewed by the author, 27 Jan1992.

---

[43] Bohm & Vigier (ref. 39).

[44] Wheeler (ref. 1), 174. John Wheeler to Stig Lundqvist, 6 Feb 1987, WP [Series II, Box Th-To, Folder Tiomno]. On Tiomno's career, see José M. F. Bassalo and Olival Freire Jr., "Wheeler, Tiomno e a física brasileira," *Revista Brasileira de Ensino de Física, 25* (2003), 426-37.

[45] Bohm, Schiller & Tiomno; Bohm & Schiller, (ref. 39).

[46] David Bohm and Jean-Pierre Vigier, "Relativistic Hydrodynamics of Rotating Fluid Masses," *PR, 109* (1958), 1882-91. " ... alors [en Brésil] Bohm et moi on a fait deux papiers, un qui a été fait de suite, qui est sorti en 1954, sur la statistique, et un deuxième qui est sorti plus tard. » Vigier, (ref. 42).

[47] The main achievements of this approach were presented in Louis de Broglie et al, "Rotator model of elementary particles considered as relativistic extended structures in Minkowski space," *PR, 129* (1963), 438-50.

[48] Bohm & Schützer, (ref. 39).

The collaboration between Bohm and Vigier, which was based initially in Brazil, reflects a certain irony typical of the Cold War. While Bohm was in Brazil for being Communist, Vigier could perhaps not have visited him in the US, for the same reason. In fact, Vigier, before becoming one of the most active spokesmen for the causal program, had already been well known as a Communist in post World War II France, and it is doubtful that he would have received an American visa to visit the US to work with Bohm. As Jessica Wang has pointed out in writing about the "age of anxiety" in American history, "in addition to refusing passports to American scientists, the State Department also restricted the entry of foreign scientists with left-wing political ties into the United States. [..] Scientists from France, where the Left was particularly strong, had an especially hard time. As much as 70 to 80 percent of visa requests from French scientists were unduly delayed or refused."[49]

For Bohm, Vigier was the most instrumental collaborator, for it was Vigier who convinced de Broglie to return to his early position of searching for a deterministic approach to quantum mechanics. A Nobel Prize winner and one of the founding fathers of the new theory, de Broglie was Bohm's most eminent ally in the defense of the causal interpretation.[50] Vigier also motivated a group of young Marxists to work on their ideas, and the Institut Henri Poincaré, in Paris, under the leadership of de Broglie and Vigier, became the main institutional base for supporters of the causal interpretation for many years.[51] After the first news from Paris, Bohm became excited with the perspectives of this collaboration ("Have you seen discussion of causal interp. of qu. theory in one of later issues of French magazine, Pensée"),[52] not only for the possibility of working on the causal interpretation, but also because Vigier's collaboration struck a very sensitive chord in Bohm's motivation to pursue the causal interpretation, a motivation related to his Marxist engagement: "I have heard from someone that in a debate on causality given in Paris, when our friend Vigier got up to defend causality, he was strongly cheered by the audience, (which contained a great many students) I would guess that many of the younger people in Europe recognize that the question of causality has important implications in politics, economy, sociology, etc."[53] However, Bohm hoped for more support from Marxists; he did not accept the distance from the causal interpretation of Philip Morrison - an American physicist well known for his Marxist positions, and he complained about the absence of support from Soviet physicists, asking why the causal interpretation had appeared in the West and not in the USSR.[54]

---

[49] Wang (ref. 7), 278).

[50] For the evolution of de Broglie's thoughts on these issues, see Louis de Broglie, "La physique quantique restera-t-elle indéterministe?" *Bulletin Soc. française de philosophie, XLVI* (1953), 135-173.

[51] Cross saw Bohm's work just as a reflection of the ideological Marxist climate of the time; thus he missed the fact that the quantum controversy continued even when that climate faded. Andrew Cross, "The crisis in physics: Dialectical materialism and quantum theory," *Social Studies of Science, 21* (1991), 735-59. A lacuna in the history of physics in the 20[th] century is the analysis of the activities of the de Broglie – Vigier group.

[52] "I have been in communication with Regner + Schatzman. They tell me about all sorts of wonderful discoveries using these new ideas, but as yet no details. I have sent them letters recently urgently asking for details." David Bohm to Miriam Yevick, 24 Dec [1952]. Ibid., *n.d.* BP.

### Brazilian funds for the causal interpretation

The support that Bohm found in Brazil for his research can also be evidenced by the funds he raised. Bohm arrived at a moment when Brazilian physics was flourishing, after Cesare Lattes's discovery, in 1947, with Cecil F. Powell and Giuseppe Occhialini, in England, of the pion in cosmic rays, and the detection by Lattes and Eugene Gardner, in the United States, of artificially produced pions. These scientific achievements resonated in Brazil, and led to an alliance between scientists, the military, businessmen, and politicians that was aimed at developing nuclear physics, and physics in general, in Brazil. This alliance led to the creation of the Centro Brasileiro de Pesquisas Físicas [CBPF] and, in the same year that Bohm arrived in Brazil, to the creation of the first federal agency exclusively dedicated to funding scientific research, the CNPq.[55] From that agency, Bohm

---

[53] David Bohm to Miriam Yevick, 5 Nov 1954. BP.

[54] "This type of inconsistency in Phil [Morrison] disturbs me. He should be helping, instead of raising irrelevant obstacles;" David Bohm to Melba Phillips, *n.d.* BP. "Then the orientation is determined strongly by the older men, such as Fock and Landau, […] It is disappointing that a society that is oriented in a new direction is still unable to have any great influence on the way in which people think and work;" idem, 18 Mar 1955. BP. "I ask myself the question 'Why in 25 years didn't someone in USSR find a materialistic interpretation of quantum theory?' […] But bad as conditions are in US etc, the only people who have thus far had the idea are myself in US, and Vigier in France." David Bohm to Miriam Yevick, 7 Jan 1952. BP.

[55] Ana M. R. Andrade, *Físicos, mésons e política: a dinâmica da ciência na sociedade* (São Paulo, 1999); idem, "The discovery of the π-meson," Helge Kragh et al, eds., *History of modern physics* (Turnhout, 2002), 313-21. Personal reminiscences from this period are in José Leite Lopes, "Cinquenta e cinco anos de física no Brasil: evocações," unpublished paper, (Rio de Janeiro, 1998), available at http://www4.prossiga.br/Lopes/. Impressions on Brazilian physics, by a contemporary visitor, are in Gordon L. Brownell, "Physics in South America," *Physics Today, 5* (July 1952), 5-12, on 11-12. It is true that, at the beginning of the 1950s, the main activities in Brazilian theoretical physics had shifted from São Paulo to Rio

received several grants to develop the causal interpretation. In 1952, the Department of Physics of USP was granted Cr\$1,246.000.00 by the CNPq, and a supplement of Cr\$528,000.00 was assured in December 1953. About 18% and 24% of those amounts, respectively, went to grants for students and visiting professors related to Bohm's activities at USP. Those funds permitted Bunge to stay in São Paulo for one year, and Schiller to have his wages supplemented for two years.[56] Besides, Bohm asked for and received Cr\$18,000.00 for the travel expenses of Schiller and his wife from the US;[57] Cr\$37,200.00 for the stay of Vigier at USP, for three months;[58] Cr\$100,000.00 for research on cosmic rays by Kurt Sitte as well as an air ticket for him and his family plus Cr\$180,000.00 to complement for one year the wages Sitte would receive from USP.[59] Bohm also won grants for the students Abrahão Zimmerman, Ruth Pereira da Silva, Paulo Roberto de Paula e Silva, and Klaus Tausk.[60] Rosenfeld, Rabi, and von Weizsäcker, physicists who interacted with Bohm but were not invited by him, also had their stays in Brazil supported by the CNPq.[61] Not all those funds went to the development of the causal interpretation, since some of them went to research on cosmic rays, a field under Bohm's responsibility at USP. Nevertheless, the board of the CNPq explicitly supported the development of the causal interpretation, as in the case when Joquim Costa Ribeiro, the Brazilian Scientific Director of the agency, presented Bohm's application for funds to support Vigier's visit. Costa Ribeiro supported the application in the following terms,[62]

> I call the attention of the Board to the interest of this subject. Prof. Bohm is today on the agenda of theoretical physics at an international level, due to his theory, which is a little revolutionary because it intends to restore in quantum mechanics the principle of determinism, which seems, in a certain way, to have been shaken de Janeiro. Nevertheless, there were close relations between the two centers, and Bohm commuted between São Paulo and Rio de Janeiro.

[56] AC [Processo 578/51]. I am grateful to Ana M. R. Andrade, and her assistants, Tatiane dos Santos and Vanessa Albuquerque, for their help in unearthing those documents.

[57] AC [Processo 572/52].

[58] AC [Processo 242/53]. It was a partial funding, insofar as Vigier also had support from France.

[59] AC [Processo 243/53]. Sitte came from Syracuse University to Brazil, after receiving an invitation from Bohm, to work on cosmic rays. Hiring Sitte was the subject of the dispute between Bohm and other members of the Physics Department, related in this paper while was commenting Bohm's discussion with von Weizsäcker. Ana M. R. Andrade, "Os raios cósmicos entre a ciência e as relações internacionais," Marcos C. Maio, ed., *Ciência, política e relações: Ensaios sobre Paulo Carneiro* (Rio de Janeiro, 2004), 215-42.

[60] AC [Processos 567/51 & 578/51].

[61] AC [Processos 1704/53, 504/53, 249/52, respectively].

[62] AC, Records of the Conselho Diretor, 139th meeting, 25 Feb 1953.

by Heisenberg's principle. Prof. Bohm seems to have found one solution to this difficulty of modern physics, trying to conciliate quantum mechanics with the rigid determinism of classical physics. I am not speaking in detailed technical terms, but summarizing the issue. Bohm's theory has given rise to a great debate in Europe and United States, and Prof. Vigier has expressed his willingness to come to Brazil, mainly to meet the team of theoretical physics and discuss the problem here. This seems to me to be a very prestigious thing to Brazil and to our scientific community.

When Bohm came to Brazil, he had not yet published his causal interpretation, and the support for his position at USP was not related to that scientific program. Once he was in Brazil, however, support from the CNPq was not independent of his program; in fact, it was conscious support for the causal interpretation.

**Bohm's uneasiness in Brazil**

If one adds to the facts previously reported the letters exchanged with Einstein, Pauli, Phillips, the debates in scientific journals with Takabayasi, Keller, Epstein, Halpern, and Freistadt, the papers by Rosenfeld, Pauli, Born, and Heisenberg, the laudatory papers by Schatzman, and Freistadt in cultural magazines, and the news Bohm had from Bohr and von Neumann, one cannot maintain that the causal interpretation passed unnoticed.[63] Noticed, however, does not mean favorably received. In fact, it had an unfavorable reception among physicists, which contributed to Bohm's uneasiness in Brazil. He was inclined to consider people intellectually meaningful inasmuch as they were receptive to the causal interpretation, and, by the same token, he did not understand people who were skeptical. It is noteworthy, for instance, as we have seen, how his high regard for Feynman was based on Feynman's receptive attitude towards the logical possibility of a causal interpretation of quantum theory, and not on Feynman's methods of renormalization, for which he was highly regarded among physicists. Bohm's hopes were not modest, "if I can succeed in my general plan, physics can be put back on a basis much nearer to common sense than it has been for a long time," and his mood oscillated depending on the reception of his ideas, or on the work he had done on them. So, we find letters saying, on one hand, "I gave two talks on the subject here, and aroused considerable enthusiasm among people like Tiomno, Schützer, and Leal-Ferreira, who are assistants […]. Tiomno has been trying to extend the results to the Dirac equation, and has shown some analogy with Einstein's field equations", and, on the other hand, " I am becoming discouraged also because I lack contact with other people, and feel that there is a general lack of interest in new ideas among physicists throughout the world."[64]

---

[63] For an analysis of Einstein's reaction, see Michel Paty, "The nature of Einstein's objections to the Copenhagen interpretation of quantum mechanics," *Foundations of Physics*, *25* (1995), 183-204; ibid. (ref. 25). Takehiko Takabayasi, "On the formulation of quantum mechanics associated with classical pictures," *Progress of Theoretical Physics, 8* (1952), 143-182; idem, "Remarks on the formulation of quantum mechanics with classical pictures and on relations between linear scalar-fields and hydrodynamical fields," *Progress of Theoretical Physics, 9* (1953), 187-222; Joseph Keller, "Bohm's interpretation of the quantum theory in terms of 'hidden' variables," *PR, 89* (1953), 1040-41; Hans Freistadt, "The crisis in physics," *Science and Society, 17* (1953, 211-37; idem, "The causal formulation of quantum mechanics of particles: The theory of de Broglie, Bohm and Takabayasi," *Nuovo Cimento, Suppl. V* (1957), 1-70; Saul Epstein, "The causal interpretation of quantum mechanics," *PR, 89* (1953), 319; Otto Halpern, "A proposed re-interpretation of quantum mechanics," *PR, 87* (1952), 389; Evry Schatzman, "Physique quantique et realité," *La pensée, 42-43* (1952), 107-22. For the view that the causal interpretation passed unnoticed, see James Cushing, *Quantum mechanics: Historical contingency and the Copenhagen hegemony* (Chicago, 1994); and Hiley (ref.4), 113.

Only by taking into account Bohm's idiosyncratic attitude towards physicists who did not share his opinion about the causal interpretation can one understand his relations with Brazilian physicists. Indeed, theoretical physicists, such as Schönberg, Leite Lopes, and even Tiomno, did not support the causal interpretation research program, insofar as one understands that program as the adoption of its philosophical premises, like the recovering of determinism, and as a rival to the Copenhagen interpretation. Tiomno collaborated with Bohm just to see what physics one could develop by using that model, not sharing Bohm's philosophical premises. Leite Lopes, a former student of Pauli's, was skeptical about the causal interpretation. Schönberg worked on the mathematical foundations of quantum theory and on the hydrodynamic model of quantum mechanics, a model close to the model Bohm and Vigier worked on, but he was against the idea of recovering a causal description in atomic phenomena. "Schönberg is 100% against the causal interpretation, especially against the idea of trying to form a conceptual image of what is happening. He believes that the true dialectical method is to seek a new form of mathematics, the more 'subtle' the better, and try to solve the crisis in physics in this way. As for explaining chance in terms of causality, he believes this to be 'reactionary' and 'undialectical.' He believes instead that the dialectical approach is to assume 'pure chance' which may propagate from level to level, but which is never explained in any way, except in terms of itself."[65] The attitude of theoretical physicists in Brazil towards Bohm's approach was not unique, but rather reflected "l'air du temps", a time in which adhesion to the Copenhagen interpretation as the only viable interpretation of quantum mechanics was widespread among physicists everywhere, and Bohm did not like this attitude.

---

[64] David Bohm to Melba Phillips, 28 June 1952; ibid., [*w.d.*], BP (C.46 – C.48). David Bohm to Hanna Loewy, 6 Oct 1953, BP (C.39).

I conclude the first part of this paper by noting that Brazil was a richer environment than that depicted by historians who based their analysis exclusively on Bohm's letters. Bohm's dissatisfaction resulted not only of the country he met but also stemmed from the confiscation of his passport as well as the unfavorable reception of his causal interpretation not only in Brazil but also in the United States and Europe. It is certain, though, that Brazil was not the best place for Bohm to fight a controversy on such a hot subject. If I take Bruno Latour's idea of "centre of calculation,"[66] or the looser dichotomy between center and periphery, the USP and the Brazilian Center for Research in Physics constituted neither such a center nor a backward physics community. They should be ranked on an intermediate level of such hierarchies, and if Bohm's Brazilian exile was a nuanced context, it needs to be balanced with other aspects to estimate its true role in the poor reception of Bohm's causal interpretation. The next part of this paper is dedicated to this task.

## 2. THE RECEPTION OF THE CAUSAL INTERPRETATION
### The reaction of the Copenhagen's supporters

The main opposition to the causal interpretation came from the old generation of supporters of the Copenhagen interpretation, such as Pauli, Heisenberg, Rosenfeld, Born, and Bohr.[67] The best description for the intellectual authority of this team, on issues related

---

[65] For Tiomno's stance, Freire Jr. (ref. 9), 95. David Bohm to Miriam Yevick, 24 Oct 1953. BP. For Schönberg's work on quantum mechanics and geometry, see Mario Schönberg, "Quantum Theory and Geometry," *Max Planck Festschrift* (Berlin, 1958), 321.

[66] A "centre of calculation" is a dynamic locus in the networks in which knowledge circulates. It is the "cycle of accumulation that allows a point to become a *centre* by acting at a distance on many points." Bruno Latour, *Science in action: How to follow scientists and engineers through society* (Cambridge, MA, 1987), 222.

[67] Rosenfeld's and Pauli's reactions will be analyzed in this paper. For Heisenberg's reactions, see Werner Heisenberg, "The Development of the Interpretation of the Quantum Theory," Wolfgang Pauli et al, eds.,

to the interpretation of quantum mechanics, is that already quoted by Max Jammer, who spoke about an "almost unchallenged monocracy of the Copenhagen school in the philosophy of quantum mechanics." Along the same lines, Mara Beller analyzed "the Copenhagen dogma" as "the rhetoric of finality and inevitability." As she wrote, "the founders and followers of the Copenhagen interpretation advocated their philosophy of physics not as a possible interpretation but as the only feasible one." My discussion of Rosenfeld's and Pauli's reactions entirely agrees with her analysis. Additionally, while analyzing commemorative practices in physics, she gave a description of the dominance of the Copenhagen interpretation, appealing for Bohr's role as a "charismatic leader," a combination of "intimidating authority and irresistible charisma," concluding that "Bohr's unprecedented authority resulted […] in an uncritical following of the Copenhagen philosophy." Interesting as Bohr's personality could be, I intend to deal with the dominance of the Copenhagen interpretation in a broader context.[68]

Pauli and Rosenfeld, were the first to react, and were the more influential. In fact, there was a division of labor between them; Pauli concentrated on the physical and epistemological aspects, while Rosenfeld dealt with the philosophical and ideological ones. Indeed, the division of labor was not planned, but it became clear when they exchanged letters while writing their papers for de Broglie's Festschrift, as Rosenfeld explained to Pauli, "*My own contribution to the anniversary volume has a different character. I deliberately put the discussion on the philosophical ground, because it seems to me that the root of evil is there rather than in physics.*" I have already discussed Pauli's early criticisms and Bohm's replies, which eventually led Pauli to recognize the logical consistency of Bohm's proposal. After Bohm's papers appeared in print, Pauli came with new criticisms, which surprised Bohm: "I am surprised that Pauli has had the nerve to publicly come out in favor of such nonsense." Bohm was infuriated, "I certainly hope that he publishes his stuff, as it is so full of inconsistencies and errors that I can attack him from several different directions at once." Pauli had criticized the fact that the causal interpretation did not preserve the symmetry between position and momentum representations, expressed in the standard formalism by the theory of unitary transformations, which was responsible for the mathematical elegance of the theory; had considered a weakness of the causal interpretation the absence of its relativist generalization; and had stated that the meaning of $\Psi$ in Bohm's model had been borrowed from the quantum theory, a criticism we had already commented. The dispute was not limited to the published papers. In a letter to Markus Fierz, Pauli used his bitter irony, writing that he was not surprised with the alliance between de Broglie and Vigier aiming to restore determinism to physics. He argued that both Catholics and Communists depended on determinism for reassuring their eschatological faiths, the former in the heaven after earthy life, the latter in the heaven still on earth. Pauli also warned "Beppo", his old friend, against sympathy with Bohm's approach. The Italian physicist Giuseppe Occhialini had worked at USP during the 1930s, and kept up his scientific collaboration with that university after the Second World War. Pauli wrote to him: "Beppo, what about South America? (For all cases I warn you for Bohm in São Paulo and his 'causal' quantum theory)."[69]

The division of labor between Pauli and Rosenfeld left the latter to the philosophical and ideological criticisms. For him, complementarity was both a direct result of experience and an indelible part of quantum theory.[70] Since complementarity implied the abandon of determinism, Rosenfeld saw the causal interpretation as a metaphysical – not scientific – attempt at dealing with quantum phenomena. According to him, "Determinism has not escaped this fate [to cease to be fertile and become obstacles to progress]; the physicist who still clings to it, who shuts his eyes to the evidence of complementarity, exchanges (whether he likes it or not) the rational attitude of the scientist for that of the metaphysician;" and he appealed for Marxist authors to reinforce his position, "The latter, as Engels aptly describes him, considers things 'in isolation, the one after the other and the one without the other,' as if they were 'fixed, rigid, given once for all.'"[71] To understand the blend of philosophy and ideology in Rosenfeld's argument, one needs to consider that besides being a physicist very sensitive to philosophical matters, that most of his work as Bohr's assistant was related to epistemological matters, that he had been engaged in Marxism since the thirties, and that Rosenfeld's Marxism was closer to Western Marxism than it was to Soviet Marxism, to use the terms introduced by Perry Anderson in order to make sense of Marxist trends in the 20th century.[72] Rosenfeld was convinced that complementarity was a dialectical achievement[73] that should be defended not only from Bohm's criticisms but also from Soviet critics who blamed it for introducing idealism in physics. As this Soviet criticism was in tune with the ideological climate in the USSR at the time, we can say that Rosenfeld was both orthodox in quantum mechanics and heterodox in Marxism.

[69] Léon Rosenfeld to Wolfgang Pauli, 20 Mar 1952, emphasis added, Pauli (ref. 25), on 587-8. Bohm to Beck [w/d], Guido Beck Papers. Beck had reported to Bohm's the content of Pauli's seminar in Paris, in 1952. The criticisms were published in Pauli's contribution to the Louis de Broglie Festschrift, see Pauli (ref. 40). Pauli to Markus Fierz, 6 Jan 1952, Pauli, (ref. 25)., 499-502; Pauli to Giuseppe Occhialini, [1951-1952]. Archivio Occhialini 5.1.14, Università degli studi, Milan. I thank Leonardo Gariboldi for calling my attention to this document.
[70] In the French version of the paper, Rosenfeld emphasized the idea of complementarity resulting from experience, but in the English version, reacting to criticisms from Born, he attenuated his stand, changing "La relation de complémentarité comme donné de l'expérience" to "Complementarity and experience."

Rosenfeld mobilized all the means he had to fight for complementarity and against the causal interpretation. He wrote to Frédéric Joliot-Curie – a Nobel prize winner and member of the French Communist Party - pushing him to take a position against French Marxist critics of complementarity;[74] advised Pauline Yates – Secretary of the "Society for cultural relations between the peoples of the British Commonwealth and the USSR" – to withdraw from *Nature* her translation of a paper from the Soviet physicist Yakov Ilich Frenkel, which was critical of complementarity;[75] took the initiative of writing to *Nature*'s editors suggesting that they not to publish a paper by Bohm entitled "A Causal and Continuous Interpretation of the Quantum Theory;"[76] and advised publishers not to translate into English one of de Broglie's books dedicated to causal interpretation.[77] Rosenfeld's stands found much support, as we can deduce from the letters he received and the fact that his first paper criticizing the causal interpretation was published in French, English, and Japanese.[78] Denis Gabor wrote, "I was much amused by the onslaught on David Bohm, with whom I had a long discussion on this subject in New York, in Sept. 51. Half a dozen of the most eminent scientists have got their knife into him. Great honour for somebody so young;"[79] and letters of support include Abraham Pais,[80] Guido Beck, Robert Cohen,[81] Eric Burhop, Vladmir Fock,[82] Jean-Louis Destouches,[83] Robert Haveman,[84] and

---

la nécessité d'en comprendre le sens exact et profond avant de se lancer dans des discussions avec des citations qui ne sont que des planages trahissant parfois leurs auteurs." Léon Rosenfeld to Frédéric Joliot-Curie, 6 Apr 1952; Joliot to Rosenfeld, 21 Apr 1952. RP. See also Michel Pinault, *Frédéric Joliot-Curie* (Paris, 2000), 508.

[75] Pauline Yates to Léon Rosenfeld, 7 Feb 1952 & 19 Feb 1952. RP.

[76] Rosenfeld succeeded, "the editors stopped work on this article." The paper had been submitted to *Nature* by Harrie S. W. Massey. *Nature*'s editors to Léon Rosenfeld, 11 Mar 1952. RP. "Also I sent a brief article to Massey with the suggestion that he publish it in *Nature*." David Bohm to Miriam Yevick, *n.d.* BP. Bohm did not keep a copy of the unpublished paper, but there is a copy of it in Louis de Broglie Papers, Archives de l'Académie des sciences, Paris.

[77] Léon Rosenfeld, "Report on L. de Broglie, La théorie de la mesure en mécanique ondulatoire." *n.d.* RP. This book had been published in 1957, Paris: Gauthier-Villars.

[78] Rosenfeld (ref. 33). The Japanese translation was published in *Kagaku, 25* (1955).

[79] Denis Gabor to Léon Rosenfeld, 7 Jan 1953. RP.

[80] "I find your piece about complementarity interesting and good. […] I could not get very excited about Bohm. Of course it doesn't do any good, but (with the exception of Parisian reactions) it also doesn't do any harm. I find that Bohm wastes his energy and that it will harm him personally a lot because he is moving into the wrong direction – but he needs to realize this himself, he is a difficult person." Abraham Pais to Léon Rosenfeld, 15 May [1952]. RP. I thank Katrien Straeten for the translation from the Dutch.

[81] "I turn to you because my own reaction to the Bohm thing and to the pilot wave revival has been quite negative, while yet I share Professor Einstein and others' uneasiness at the orthodox situation." Robert Cohen to Léon Rosenfeld, 31 Jul 1953. RP.

[82] "Je voudrais aussi discuter avec vous les questions d'interprétation de la mécanique quantique et surtout les causes et les effects de la 'maladie Bohm-Vigier', assez repandue, hélas." Vladmir Fock to Léon Rosenfeld, 7 Apr 1956. RP. For Fock's criticism of Bohm's views, see Vladmir Fock, "On the interpretation of quantum theory," *Czechosl Journ Phys, 7* (1957), 643-56.

[83] Jean-Louis Destouches to Léon Rosenfeld, 19 Dec 1951. RP. This letter is a fair description of de Broglie's hesitations before his conversion to causal interpretation. It also describes the French philosophical context in which the causal interpretation was well received: "L'indéterminisme quantique et les conceptions de Bohr et Heisenberg n'ont jamais été admises en France sauf par M. Louis de Broglie et ses élèves. […] Les jeunes gens ont accueilli avec enthousiasme le travail de Bohm qui correspond à toutes les tendances philosophiques

Adolf Grünbaum.[85] Some of them, however, did not accept all the incisiveness of
Rosenfeld's rhetoric, as was the case with Burhop and Beck.[86] Later, when Rosenfeld
published a review of Bohm's "Causality and Chance in Modern Physics," the same style
cost Rosenfeld public disagreement with Lancelot L. Whyte.[87]

Rosenfeld, Pauli, Heisenberg, and Born successfully built a common front against
the causal interpretation, but they needed to deal, mainly privately, with their divergences.
Whereas Rosenfeld kept up a lasting public debate with Heisenberg until 1970, criticizing
his leaning towards idealism, Pauli and Born privately criticized Rosenfeld's mixture of
Marxism with complementarity. As part of the debate, Max Born wrote and sent to
Rosenfeld a 10-page typed text in which he argued that dialectical materialism could not
be corroborated by reference to just one achievement of contemporary science. Ultimately,
Born abandoned the idea of publishing the text having seen the beginning of a détente
between West and East in the late 1950s. Wolfgand Pauli used his famous ironic and bitter
correspondence style to hit Rosenfeld. When editing a volume in honor of Bohr, he wrote
to Heisenberg saying that he had managed to prevent Rosenfeld, whom he labeled
Rosenfeld "•*BohrxTrotzky*," from adorning his paper with banalities on Materialism.[88]

### The label of "philosophical controversy"

---

qui les animent : réalisme thomiste, déterminisme marxiste, rationalisme cartésien. Je suis donc maintenant à
peu près le seul ici à soutenir encore l'interprétation quantique de Bohr."
[84] "I read with great interest your paper and I am glad seeing that our ideas are, in their essential aspects, in
agreement." Léon Rosenfeld to Robert Haveman, 7 Oct 1957; Haveman to Rosenfeld, 13 Sep 1957. RP.
[85] Adolf Grünbaum to Léon Rosenfeld, 1 Feb 1956, 20 Apr 1957, 3 Oct 1957; Rosenfeld to Grünbaum, 14
Feb 1956, 21 May 1957, 11 Dec 1957. RP.
[86] "Incidentally the only other comment I would offer on your article was I thought perhaps you were a little
cruel to Bohm. Do you think you could spare the time to write to him? He is a young Marxist […] being
victimized for his political views in the US." Burhop was organizing a meeting among Rosenfeld, John
Bernal, Maurice Levy, Maurice Cornforth, and Cecil Powell, to discuss Rosenfeld's article. Eric Burhop to
Léon Rosenfeld, 5 May 1952. RP. I discussed in Part I Beck's stands.
[87] Lancelot Whyte to Léon Rosenfeld, 8 Apr 1958; 14 Mar 1958; 22 Mar 1958; 27 June 1958; Rosenfeld to
Whyte, 17 Mar 1958, RP. Rosenfeld to Whyte, 28 May 1958, is in L. L. Whyte Papers, Boston University,
Department of Special Collections. Léon Rosenfeld, "Physics and metaphysics," *Nature*, *181* (1958), 658;
Lancelot Whyte, "The scope of quantum mechanics," *The British Journal for the Philosophy of Science*, *9*
(1958), 133-4.
[88] Léon Rosenfeld, "Heisenberg, physics and philosophy," *Nature 186* (1960), 1960, 830-1; ibid., "Berkeley
*redivivus*," *Nature 228* (1970), 479. Olival Freire Jr., (ref. 70). Pauli to Heisenberg, 13 May 1954; Pauli to
Rosenfeld, 28 Sep 1954, *Wolfgang Pauli – Scientific Correspondence,* Vol. IV, Part II (1953-1954), [ed. by
Karl von Meyenn], (Berlin, 1999), 620-1, and 769.

Now I shall narrow my analysis to the discourse shared by Copenhagen's supporters while criticizing the causal interpretation; the reaction of causal interpretation's supporters to that discourse; and the social, and especially professional, consequences of such a dispute. Pauli considered that insofar as Bohm's approach had "never any effects on observable phenomena, neither directly nor indirectly, […] the artificial asymmetry introduced in the treatment of the two variables of a canonically conjugated pair characterizes this form of theory as *artificial metaphysics*." [If the] "new parameters could give rise to empirically visible effects, […] they will be in disagreement with the general character of our experiences, [and] in this case this type of theories loses its physical sense."[89] Rosenfeld, as we have already seen, "deliberately put the discussion on the philosophical ground, because it [seemed to him] that the root of evil is there rather than in physics."[90] In the published paper, he used softer words but with the same content: "I intentionally confine the debate to the field of epistemology, for the crucial issue is one of logic, not of physics." About physics, he conceded, "Bohm's argument is very cleverly contrived. One would look in vain for any weakness in its formal construction."[91] Heisenberg termed causal interpretation as an "ideological" attempt.[92] Niels Bohr and Max Born, quite apart from their reactions to the causal interpretation, always emphasized the epistemological nature of some choices related to interpreting quantum formalism. If metaphysics, philosophy, epistemology, and ideology were the terms, how could a physicist frame the controversy between causal or complementarity interpretation in the early 1950s? They saw the controversy, in the best case, as an expression of a strictly philosophical dispute concerning ontology (the constitution of the real physical objects as waves or/and particles) and epistemology (the status of determinism in physical theories, the completeness of theories, the role of the space-time description). In the worst case, 'metaphysical' was used as an adjective defining disputes without implications for the development of physics. Even physicists who tried to present the controversy impartially shared this view. Albert Messiah's case is exemplar. In his very influential textbook, published originally in 1958, he wrote, "the controversy has finally reached a point where it can no longer be decided by any further experimental observations; it henceforth belongs to the philosophy of science rather than to the domain of physical science proper."[93] A similar example is Fritz Bopp's statement, during a conference dedicated in 1957 to foundational problems in quantum mechanics: "…what we have done today was predicting the possible development of physics – we were not doing physics but metaphysics."[94]

---

[89] Pauli (ref. 40). Emphasis added.
[90] Ref. 69.
[91] Rosenfeld (ref. 33).
[92] Heisenberg, *Physics* (ref. 67), 133.

　　　　To analyze how the supporters of the causal interpretation reacted to the philosophical controversy diagnosis, let us recall what result Bohm and his collaborators obtained. Their main achievement was the empirical equivalence with nonrelativistic quantum mechanics. Bohm and his colleagues consistently searched in the 1950s for results derived from the causal interpretation approach which were able to contrast such an approach with the usual quantum mechanics. He announced throughout the 1950s that he was developing a satisfactory relativistic generalization of his approach. Neither of those promises was fulfilled,[95] and this failure discouraged some of his supporters.[96] The absence of new results reinforced the "philosophical" label of the dispute, but Bohm and his colleagues did not manage to weaken it. The original paper had had the technical title of "hidden" variables; afterwards, he shifted to name his own approach using the more philosophical term "causal interpretation." His 1957 book was meaningfully entitled, from a philosophical point of view, "Causality and Chance in Modern Physics." Vigier, acting in the more liberal context of France, presented Bohm's and his own work as an "illustration of dialectical materialism."[97] I believe that Bohm and his collaborators trapped themselves by accepting the "philosophical controversy" diagnosis.[98] To develop the causal program, they needed to have people, mainly young physicists, working on it, and it could not be easy, if not impossible, to introduce in the agenda of research in physics a theme considered part of a philosophical controversy without creative implications for physics. The label of "philosophical controversy" contributed to keeping away new entrants into physics, because, after all, a career in physics is not a career in philosophy. Surely, the ideological label, as used by Vigier, appealed to young Marxist physicists, but this appeal, as effective as it was in the 1950s, was not enough to sustain such a research program. I will later consider other difficulties related to introducing the causal program in the agenda of physics research, obstacles concerning how quantum mechanics was being taught.

---

[93] Albert Messiah, *Quantum mechanics* (Amsterdam, 1964), vol I, 48. However, he did not please the hard core of the supporters of the Copenhagen interpretation. Rosenfeld wrote to him praising the book, but in disagreement with his diagnosis of the controversy. For Rosenfeld, "Ce n'est pas en effet d'expérience, mais bien de simple logique qu'il s'agit ici." Léon Rosenfeld to Albert Messiah, 16 Jan 1959. RP.

[94] Bopp, in Stephan Körner, ed., *Observation and interpretation in the philosophy of physics, with special reference to quantum mechanics* (New York, 1957), 51. By the way, Bopp was working on another alternative interpretation, the so-called "stochastic interpretation."

[95] My previous writings on Bohm's case overestimated these aspects, since they were not put in the broader context that I am discussing in this paper.

[96] Two examples are Mario Bunge and Philippe Leruste. "However, as time went by and no new predictions came out of the new formulation, I started to have doubts. Then, in 1964, when I started working on the axiomatization of NRQM for my *Foundations of physics* (Springer, 1967), I realized that Bohm's was not a valuable addition to standard QM and that the solution to his (and de Broglie's and Einstein's) problems lay elsewhere, namely in a realistic reinterpretation of standard QM." Mario Bunge to the author, 1 Nov 1996 & 12 Feb 1997. Ph. Leruste to the author, 27 Jan 1992.

[97] Jean-Pierre Vigier, "Quelques problèmes physiques posés par les theses de Lénine," *La pensée, 57* (1954), 60-66.

### The working of a monocracy - quantum mechanics training and research agenda

Jammer's description of the "almost unchallenged monocracy of the Copenhagen school in the philosophy of quantum mechanics" does not tell us how physicists adhered to such a school, or how they understood complementarity, the conceptual core of that school. John Heilbron, discussing the first missionaries of the Copenhagen doctrine provides some hints on these questions. He suggests that, beyond Bohr's close circle (Heisenberg, Pauli, Jordan, Born, Rosenfeld) and their brilliant opponents (Einstein, Schrödinger), physicists did not consciously adhere to complementarity or criticize it, but rather used the quantum machinery to scrutinize the microscopic world. Heilbron also suggests that the philosophical flavor of Bohr's views on the interpretation of quantum mechanics was responsible for American and British indifference to complementarity.[99] Sam Schweber added two American peculiarities, both of them hostile to the idea of philosophizing on issues of quantum mechanics, namely the placing of theoretical and experimental physicists in the same departments, reinforcing experimentation and application and American trends toward pragmatism.[100] Analyzing how American physicists reacted in the 1920s to the philosophical problems of quantum theory, Nancy Cartwright argued: "Americans in general had little anxiety about the metaphysical implications of the quantum theory; and their attitude was entirely rational given the operationalist-pragmatist-style philosophy that a good many of them shared."[101] After World War II, for reasons related to the post-war context, American physics watched the reinforcement of such trends; as argued by David Kaiser, "the pedagogical requirements entailed by the sudden exponential growth in graduate student numbers during the Cold War reinforced a particular instrumentalist approach to physics." In this context, Kaiser continues, "epistemological musings or the striving for ultimate theoretical foundations – never a strong interest among American physicists even before the war – fell beyond the pale for the postwar generation and their advisors."[102]

---

[98] Before the publication of his original paper, Bohm was conscious of the risk of such a label, but apparently did not work to prevent it. "What I am afraid of is that the big shots will treat the article with a conspiracy of silence; perhaps implying privately to the smaller shots that while there is nothing demonstrably illogical about the article, it really is just a philosophical point, of no practical interest." David Bohm to Miriam Yevick [1952]. BP.

[99] John Heilbron, "The earliest missionaries of the Copenhagen spirit," Peter Galison et al, eds., *Science and society: The history of modern physical science in the twentieth century,* (New York, 2001), vol. 4, 295-330.

If one considers that for European physicists, such as Bohr and Pauli, epistemological considerations were integral to their way of doing physics, one could get a hint of the complexity of the diffusion of the quantum mechanics, which has become universal by circulating through such different intellectual and professional contexts. As a matter of fact, the textbooks in which physicists learned quantum mechanics until the 1950s did not "reflect much concern at all about the interpretation of the theory."[103] According to Helge Kragh, "most textbook authors, even if sympathetic to Bohr's ideas, found it difficult to include and justify a section on complementarity. Among forty-three textbooks on quantum mechanics published between 1928 and 1937, forty included a treatment of the uncertainty principle; only eight of them mentioned the complementarity principle."[104] Bohr's epistemological writings were circulated in papers presented in scientific meetings, and put together in anthologies, which are vehicles quite different from textbooks; physicists are formed, as remarked by Thomas Kuhn, mainly via textbooks.[105] The lack of complementarity's lessons in the training of physicists was acutely felt only when it was challenged by the causal interpretation. Rosenfeld expressed his worries very clearly when he wrote "there is not a single textbook of quantum mechanics in any language in which the principles of this fundamental discipline are adequately treated, with proper consideration of the role of measurements to define the use of classical concepts in the quantal description." At the same time, he was consciously helpless, as "there is thus most obviously an urgent need for a good elementary treatise […]. But it will be extremely difficult to find an author for such a book: those who have the competence to write it are too busy with other problems."[106] Doubtless, the preferred author for Rosenfeld, and who never wrote such a book, was Niels Bohr,[107]

> There is great interest in the topic among chemists and biologists, but there is no book that one can refer them to and that could protect them from the confusion created by Bohm, Landé, and other dilettantes. I will now do my bit here in Manchester by giving a lecture for chemists and biologists; but nothing can replace the book that *you* must write.

---

[100] Sam Schweber, "The empiricist temper regnant: Theoretical physics in the United States 1920-1950," *HSPS, 17(1)* (1986), 55-98.
[101] Nancy Cartwright, "Philosophical problems of quantum theory: The response of American physicists." Lorenz Krüger et al, eds., *The probabilistic revolution* (Cambridge, MA, 1987), vol 2, 407-435. For an analysis along the same lines, see Katherine R. Sopka, *Quantum physics in America 1920-1935* (New York, 1980), 3.67-3.69. Assmus suggests a more prosaic explanation for this little anxiety, since his thesis is: "aspiring quantum scientists chose the field of molecular structure in which to make their mark, avoiding the competitive field of atomic physics, which by the 1920s has become the cutting-edge of European physics." Philosophical problems arose mainly in atomic physics, once they were just implicit in molecular physics. Alexi J. Assmus, "The Americanization of molecular physics," *HSPS, 23(1)* (1992), 1-34. I thank David Kaiser for calling my attention to this paper.
[102] Kaiser (ref. 4, *HSPS*), 154-6.
[103] Jagdish Mehra and Helmut Rechenberg, *The historical development of quantum theory* New York, 2001), vol. 6, part 2, 1194.

If the previous account seems plausible, it implies that Jammer's description hides a social and intellectual division of labor among physicists. The monocracy of the Copenhagen school referred to two types of physicists. A few of them were involved with foundational problems, the extension of quantum mechanics to new phenomena, and its applications to old and new problems; the others were involved just with extension and applications because they believed that the foundational problems were well solved by the founding fathers of quantum mechanics. This context was very adverse to Bohm's ideas.

---

[104] Helge Kragh, *Quantum generations: A history of physics in the twentieth century* (Princeton, 1999), 211.
[105] Thomas Kuhn, *The structure of the scientific revolutions* (Chicago, 1970).
[106] Léon Rosenfeld, "Report on L. de Broglie," (ref. 77). The report follows: "The nearest to a really good treatment is found in Landau and Lifschitz's outstanding treatise: but it is too short and not explicit enough to be a real help to the student. The only books which are purposely devoted to an exposition of the principles are v. Neumann's aforementioned treatise and a little book by Heisenberg: the first is (as stated above) misleading in several respects, the second is too sketchy and on the subject of measurements it even contains serious errors (however surprising this may appear, the author being one of the founders of the theory). As to Bohr's authoritative article, it is in fact only accessible to fully trained specialists and too difficult to serve as an introduction into this question."
[107] Léon Rosenfeld to Niels Bohr, 14 Jan 1957. Bohr Scientific Correspondence (31), Archives for the History of Quantum Physics, American Philosophical Society.

People to do the work, physicists and physics students, were needed not only for developing the causal interpretation but also for the whole ensemble of issues related to foundations of quantum mechanics. However, it was not feasible for people to begin a career in, or to shift their interests to, a subject that was beyond both the training and the agenda of research in physics. It was not by chance that many of those who overcame such a barrier had interests beyond the field of physics, as in the case of the young French Marxist physicists. Keeping this context in mind, one can realize how Bohm was uneasy with physicists who were interested in another agenda of research, like Feynman, and pleased by the news he had from France. It is remarkable, finally, that the same absence of foundational issues in the textbooks, which in the short term ran against the critics of the complementarity interpretation, ran against complementarity itself, in the medium term. As time went by and the number of people interested in such issues grew, especially in the 1960s, the absence of complementary in the training of physicists fertilized the ground for its critics.[108]

## A "field of struggles to conserve or transform this field of forces"[109]

Jammer's reference to a monocracy elicits Pierre Bourdieu's idea of scientific field with its unequal distribution of capital, symbolic but no less effective. Indeed, as early as 1977, Pinch, in a well argued paper, used Bourdieu's sociological framework of "scientific field" to argue that Bohm had successfully followed a "succession strategy" before 1952, i.e. accumulating symbolic capital, and had then switched to a "subversion strategy" with the publication of his heterodox paper on "hidden variables," in 1952. Pinch doubted whether the strategy of publishing an "interpretation" had been the most adequate strategy to challenge the quantum theory and change that scientific field, but he saw the reaction against Bohm's causal interpretation just as a "conservationist strategy" of the "elite" of quantum physics. As he wrote,[110]

> the attacks on Bohm by the quantum elite can be regarded as part of what Bourdieu calls the conservationist strategy to be followed by the elite to ensure continual return on their investments. Bohm, by advocating a heterodox interpretation, was challenging the elite's authority by questioning the legitimacy of their previous investments in the interpretation of quantum theory. The official-history mode articulation of von Neumann's proof can be regarded then as an attempt to maintain a particular authority structure.

---

[108] Freire (ref. 3).
[109] For Bourdieu, "the scientific field, like other fields, is a structured field of forces, and also a field of struggles to conserve or transform this field of forces." Pierre Bourdieu, *Science of science and reflexivity* (Cambridge, 2004), 33, transl. By R. Nice. For a critical review of Bourdieu's stance on the new sociology of science, Helene Mialet, "The 'righteous wrath' of Pierre Bourdieu," *Social Studies of Science*, *33/4* (2003), 613-621.

Pinch's paper remains, in its essence, an appealing approach to understanding the poor reception of Bohm's causal interpretation, and the bulk of archival documents unearthed since then seems to confirm this. While he adequately concluded that criticisms like Pauli's and Rosenfeld's "are criticisms which are along the metaphysical dimension of scientific activity and do not involve the construction of a specific cognitive object onto which the dispute could crystallize," his analysis was focused on the symbolic role of von Neumann's proof. Indeed, its focus on von Neumann's proof may not afford a comprehensive view of the monopolies Bohm was challenging with his "subversion strategy."[111] von Neumann's proof was just part of what I am calling, after Jammer, the monocracy of the Copenhagen school. That proof did not play a role, for instance in the arguments by Rosenfeld, Pauli, and Heisenberg, nor did it play a role in the skepticism of physicists who were interested in other issues of research, like Feynman, Beck, and Leite Lopes. Incidentally, from what we know nowadays, von Neumann himself was less active in criticizing Bohm than were his European colleagues more identified with the Copenhagen interpretation and closer to Bohr. Keeping Bourdieu's frame of scientific field, one can state that the investment in the idea that complementarity had already solved the foundational problems of quantum mechanics was bigger than in von Neumann's proof, and for this reason it was defended with more determination.

Von Neumann did not publish any paper criticizing the causal interpretation, and the few leads which appeared in Bohm's correspondence suggested that he was more open-

---

[110] Trevor Pinch, "What does a proof do if it does not prove? A study of the social conditions and metaphysical divisions leading to David Bohm and John von Neumann failing to communicate in quantum physics," Everett Mendelsohn et al, eds., *The social production of scientific knowledge* (Dordrecht, 1977), 171-216, on 206. I thank Joan Bromberg for calling my attention to this paper.
[111] Pinch, ibid., 183.

minded towards Bohm's work than were physicists like Pauli, Rosenfeld, Bohr, and Heisenberg. Indeed, Bohm knew that "von Neumann thinks my work correct, and even 'elegant,' but he expects difficulties in extending it to spin."[112] It is thus an intriguing puzzle in the history of science to know the grounds in which von Neumann reacted to Bohm's proposal. The puzzle is not one of easy solution due to the absence of clear-cut documents. It is sure that in the 1950s he paid attention to these issues because, in spite of being completely busy with computer and military research. He did carefully revise the English translation of his 1932 "Mathematische Grundlagen der Quantenmechanik," in which that proof appeared, paying great attention to the philosophical aspects of the subject. To his publisher, he explained the delays with his revision of the translation, "… the text had to be extensively rewritten, because a literal translation from German to English is entirely out of question in the field of this book. The subject-matter is partly physical-mathematical, partly, however, a very involved conceptual critique of the logical foundations of various disciplines […]. This philosophical-epistemological discussion has to be continuously tied in and quite critically synchronised with the parallel mathematical-physical discussion." Indeed, there was a true tale behind the American publication of that book. von Neumann began it in 1945, and for ten years he faced problems with the copyrights (which were vested by the United States during the war), with finding adequate mathematical types, and with the translation. R.T. Beyer, suggested by Dover, did the first translation, but von Neumann carefully rewrote all of it. Eventually, it was published in 1955, by Princeton University Press, once Dover gave it up due to lengthy delays. A convincing solution for the puzzle on von Neumann's reaction to Bohm's hidden variables was recently suggested by Michael Stöltzner, which was based on a philosophical reconstruction of von Neumann's methods. Stöltzner argues that von Neumann's criteria for the success of physical theory were,[113]

---

[112] By contrast, in the same letter, Bohm says that "the elder Bohr [Niels Bohr] didn't say much to Art[hur] Wightman, but told him he thought it 'very foolish.'" David Bohm to Melba Phillips, *n.d.* BP (C.46 – C.48). The same comment on von Neumann's reaction can be found in David Bohm to Wolfgang Pauli, [Oct 1951], in Pauli (ref. 25), 389-94.
[113] John von Neumann to H. Cirker, [President of Dover Pub], 3 Oct 1949. John von Neumann Papers [Box 27, Folder 8], Library of Congress, Washington. John von Neumann, *Mathematical foundations of quantum mechanics* (Princeton, 1955).Michael Stöltzner, M. 1999. "What John von Neumann thought of the Bohm interpretation," Daniel Greenberger et al, eds., *Epistemological and experimental perspectives on quantum physics* (Dordrecht, 1999), 257-262, 260. On Bohm's program he wrote: "First, one is faced with two equations instead of one, without gaining new empirical predictions. […] Due to its inherent non-locality, the

empirical adequacy in the narrow sense, […] simplicity of the description scheme, heterogeneity of the material described by it, and fertility for further developments; [and concludes that], as to these aesthetic criteria, the Bohm program performs rather poorly, […] von Neumann could accept Bohm's proposal as an interesting model, but not as a promising interpretation.

Rather, and more important to our attempt of contrasting von Neumann's reaction with those by Pauli and Rosenfeld, Stöltzner was able to recover scattered fragments in von Neumann's papers – like this, in 1955, "there have been in the last few years some interesting attempts to revive other interpretation" – which permitted him to state, "taking into account how fiercely Pauli rebutted the Bohm interpretation, these lines presumably represent the most conciliatory reaction to it among the fathers of the 'Copenhagen interpretation.'" Thus it is time to remark that Bohm and von Neumann might have had free and informal discussions about the subject if Bohm had remained at Princeton. That opportunity was lost with Bohm's exile, and it can be attributed to the effects of McCarthysm and Bohm's exile on his research program.

Until now, I have been using freely terms like "Copenhagen interpretation," "complementarity interpretation," and "orthodox interpretation," and equating them to Bohr's thoughts on quantum mechanics. However, this is problematical, as argued by Catherine Chevalley, who was interested in a question that I am not dealing with in this paper, i.e. why Bohr is usually considered an obscure thinker, but which presents a valuable consideration for our discussion. Chevalley convincingly argued that one could not understand Bohr's reflections independent of their context, and that this context was related, on one hand, to the "history of atomic physics," and on the other hand, to the "history of a philosophical tradition widely different in content from either logical positivism of *Lebensphilosophie*." This last tradition would be more understandable if one takes into account "the very precise lexicon of post-Kantian epistemology." For what interests us in this paper, Chevalley argued, "the term 'Copenhagen interpretation' appear[ed] only in the mid-1950's in the context of hidden-variables and Marxist materialism," which led her to consider that Bohr's thoughts were distorted and assimilated

Bohm interpretation is hardly fertile in the view of the successes of local quantum field theories in elementary particle physics. [...] Bohm's interpretation does not contain any new constant that would represent some new subquantum physics." On von Neumann and foundations of quantum mechanics, see Miklós Rédei and Michael Stöltzner, *John von Neumann and the foundations of quantum mechanics* (Dordrecht, 2001).

to a term whose exact content has been the object of wide disagreement. Chevalley also remarked that, among the critics, the first to use the term "Copenhagen school" had been the Soviet physicist Blokhinzev and, among the supporters, the first had been Heisenberg.[114] Independent evidence, which reinforces Chevalley's point, is that when Heisenberg did, he was criticized by Rosenfeld, who argued that such a label could lead people to admit the existence of other interpretations. Heisenberg conceded that Rosenfeld was right,[115]

> I avow that the term 'Copenhagen interpretation' is not happy since it could suggest that there are other interpretations, like Bohm assumes. We agree, of course, that the other interpretations are nonsense, and I believe that this is clear in my book [*Physics and Philosophy*], and in previous papers. Anyway, I cannot now, unfortunately, change the book since the printing began enough time ago.

For the purposes of this paper, I can say that the very term "Copenhagen school", which Jammer described as having the quasi monopoly of the truth on the issues related to the foundations of quantum mechanics, was rather the result of the battles of the 1950s than the result of a natural and continuous evolution since the 1920s, battles that affected the causal interpretation, isolating it among physicists, but that also affected the followers of the thinking of the founding fathers of quantum mechanics, distorting their views.

### Some comparative perspectives

Before arriving at my conclusion, it is helpful to raise two comparative essays. The first is related to the role of the McCarthyist climate in the reception of the causal interpretation in different national contexts. The second is a comparison between the receptions, in the same context, of two alternative interpretations. When we compare the United States and Europe concerning the reception of the causal interpretation, it becomes clear that main characters were located in Europe: Pauli, Rosenfeld, Born, Heisenberg, and Fock, who wrote and acted against it, worked in Switzerland, England, Germany, and the USSR, and de Broglie and Vigier supported it in France. In America, the main reaction came from Einstein, who was a critic both of complementarity and the causal interpretation. Einstein also openly criticized McCarthyism.[116] Thus the fate of the causal interpretation, in the 1950s, was decided by actors not influenced by McCarthyism, or even in opposition to the anticommunist hysteria in the US. We have also seen that Bohm's status as a Communist and a victim of McCarthyism did not work against him or his proposal in Brazil. Thus, we can arrive at the conclusion that McCarthyism was not an influential factor in the real battles between Copenhagen's supporters and opponents.

---

[114] Catherine Chevalley, "Why do we find Bohr obscure?", Greenberger (ref. 120), 59-73.
[115] Werner Heisenberg to Léon Rosenfeld, 16 Apr 1958. RP.

It remains an open question whether the McCarthyist climate prevented American physicists from discussing the causal interpretation.[117] This possibility remains plausible, but the available documentary evidence does not afford a clear-cut answer. However, in order to settle how the causal interpretation was received among American physicists, we still need to examine other factor. In fact, Russel Olwell, who considered that Bohm's persona as somebody tainted by McCarthyism was an obstacle to the reception of his ideas in the United States, produced a more elaborate analysis because besides the idea of a politically tainted Bohm, he appeals to some features of American physics after World War II, such as military funding (following Forman) and pragmatic tendencies (following Schweber), to argue that US physics was not receptive to any work that put in question the foundations of theoretical physics.[118] Given the context of American physics, previously discussed, including its features related to the training of physicists, one can conjecture that, "whereas European physicists might indeed have neglected Bohm's work because most of them had already made up their mind in favor of the Copenhagen interpretation, most in the US didn't even think they had to make a choice (Copenhagen vs Bohm)."[119]

A comprehensive answer to the role played by ideological and political factors in the quantum controversy needs to consider at least one other effect, because history of Cold War times, like history tout court, must be comprehensive. As Graham once remarked, "In the very period when Soviet politicians were finding bourgeois idealism lurking in the minds of Soviet scientists, many American politicians were convinced that the State Department was infested with Communists." In fact, Bohm's persona as a Communist victim of McCarthyism was one of the effects of the Cold War climate on the reception of the causal interpretation, the other being the support it received from the young Marxist physicists, mainly located in France around Vigier, who saw research on the causal interpretation as part of the ideological battles of the times. I have argued elsewhere that the role played by Marxist criticism against the complementarity interpretation in the USSR and in the West was more influential than we have already recognized. Thus, the incidence of politics and ideology on the controversy about the interpretation of quantum mechanics still deserves more historical research.[120]

---

[116] Abraham Pais, *Einstein lived here* (New York, 1994); Fred Jerome, *The Einstein file* (New York, 2002).
[117] For authors who answered this question positively, see Olwell; Hiley; and Peat; all on (ref. 2).
[118] Olwell (ref. 2); Schweber (ref. 100); Paul Forman, "Behind quantum electronics: National security as basis for physical research in the United States, 1940-1960," *HSPS, 18(1)* (1987), 149-229.
[119] David Kaiser, e-mail to the author, 29 Nov 2004.

The second comparative study concerns the saga of Hugh Everett III and his dissertation written, in the second half of the 1950s, under John A. Wheeler, at Princeton. Like Bohm, Everett produced an alternative interpretation of quantum mechanics which was equally poorly received in its first ten years and revived after the late 1960s. Differently from Bohm, however, Everett did not have to face McCarthyism or exile. I will summarize the Everett's dissertation, the obstacles it faced, and then compare Everett's and Bohm's case.[121]

Everett's motivation for suggesting a new interpretation for quantum theory was related to the challenge of quantizing general relativity, which was then and is still today an unsolved problem. He remarked that one cannot deal through quantum mechanics with the idea of a closed universe, a concept that is essential for cosmologists, since "the whole interpretive scheme of that formalism [quantum theory] rests upon the notion of external observation." Indeed, Bohr's complementarity also strongly relies on the assumption that you need to use classical concepts for describing and communicating the results inscribed in measurement devices, while the system under study is treated according quantum mechanics. And yet, von Neumann's presentation of quantum theory introduced the axiomatic distinction between two processes of evolution of the quantum states: the first is discontinuous and not ruled by the Schrödinger equation, and happens during the observations; the second is the deterministic change of an isolated system, governed by the Schrödinger equation, which is valid in the absence of measurements. Everett's strategy was to dispense with the first of von Neumann's processes and to push to its ultimate consequences a quantum treatment based exclusively on the second process. Everett considered a device measurement as just a subsystem of larger systems and treated such subsystems according to quantum mechanics. This argument was in line with von Neumann's mathematical approach but far from Bohr's insistence that device measurements should be treated according to classical physics. Everett's *tour de force* was to attribute physical reality to the picture of an ever branching universe, each branch being the state of a subsystem plus the related state of the whole system, at a moment immediately after each physical interaction. Even though this scheme is far from our intuition, it is not logically inconsistent, "since all the separate elements of a superposition individually obey the wave equation with complete indifference to the presence or absence ('actually or not) of any other elements." Our common-sense intuition is preserved because "this total lack of effect of one branch on another also implies that no observer will ever be aware of any 'splitting' process."[122]

---

[120] Loren Graham, *Science and philosophy in the Soviet Union* (New York, 1972), 19; Olival Freire Jr., "Marxism and quantum controversy: responding to Max Jammer's question," unpublished paper, conference on "Intelligentsia: Russian and Soviet science," University of Georgia, Oct 2004.

[121] I have presented elsewhere what I called "the many lives of Everett's interpretation," contrasting the obstacles faced by Everett while doing his dissertation with the renaissance of his ideas ten years later; Freire (ref. 3).

Everett's 'relative state' formulation was not as heretical Bohm's causal interpretation was. He considered that "it can be said to form a *metatheory* for the standard theory," and described its advantages in dealing with some rather formal issues, such as "imperfect observations and approximate measurement" as well as in approaching a problem more central to physics, that is, its possible "fruitful framework for the quantization of general relativity." Nonetheless, less heresy is still heresy and he made no qualms about stating his distance from Bohr's epistemological considerations on quantum mechanics: "The particular difficulties with quantum mechanics that are discussed in my paper have mostly to do with the more common (at least in this country) form of quantum theory, as expressed, for example, by von Neumann, and not so much with the Bohr (Copenhagen) interpretation. The Bohr interpretation is to me even more unsatisfactory, and on quite different grounds." Because Everett was a reader of Bohm's 1951 textbook and his scheme shared Bohm's realistic commitment, one can say, in short, that he put together Bohm's realism and von Neumann's quantum treatment of measurement devices, and therefore did not accept Bohr's complementarity view about measurement. Everett's dissertation left Wheeler, his adviser, at a crossroads. He early approved the physico-mathematical scheme, "the correlation [paper] seems to me practically ready to publish," but he disliked Everett's epistemological considerations, which included a whole section on the different interpretations of quantum mechanics: "I am frankly bashful about showing it to Bohr in its present form, valuable and important as I consider it to be; because of parts subject to mystical misinterpretations by too many unskilled readers." In fact, Wheeler, as a Bohrian, could not accept Everett's rejection of against complementarity.[123]

---

[122] Hugh Everett III, " 'Relative State' formulation of quantum mechanics," *Reviews of Modern Physics*, 29 (1957), 454-62.

Paralyzed at the crossroads, Wheeler tried to follow both roads at the same time. He had the idea, which one could retrospectively call wishful thinking, of convincing Bohr of the value of Everett's approach and persuading Everett to remove the epistemological considerations from his dissertation. Eventually, Wheeler pleased neither Greeks nor Trojans, as we will see. His plans Wheeler's plans were far from modest. He wanted to publish Everett's dissertation in full at the Danish Academy of Science as a way of legitimizing it among the supporters of complementarity. "Since the strongest present opposition to some parts of it [Everett's dissertation] comes from Bohr, I feel that acceptance in the Danish Academy would be the best public proof of having passed the necessary tests."[124] In 1956, with a draft of the dissertation in his luggage, Wheeler went to Copenhagen to review it with Bohr. The discussions there included Aage Petersen and Alexander Stern, besides Wheeler and Bohr. It is worth remarking that in the 1950s Bohr's influence on the subject of the interpretation of quantum mechanics was so great that a Princeton dissertation was being reviewed in Copenhagen before being judged at Princeton University. Bohr could not, and did not, accept Everett's ideas on the epistemological considerations about observation in quantum mechanics. The temperature of the debates was so high that Stern reported Everett's point as theology, which led Wheeler to reply, "if it is a theological statement to postulate the 'universal wave function,' it is also a theological statement to refuse to entertain the postulate." The full content of the Copenhagenners's judgment survived recorded in a letter by Rosenfeld, who was not in Copenhagen but closely followed all epistemological matters related to quantum mechanics:[125]

---

[123] Hugh Everett to Aage Petersen, 31 May 1957. WP, Series I, Box Di - Fermi Award #1, Folder Everett. For the first version of the dissertation, Hugh Everett, "The theory of the universal wave function," Bryce DeWitt and Neill Graham, eds., *The many-worlds interpretation of quantum mechanics* (Princeton, 1973), 3-140; John Wheeler to Everett, 21 Sep 1955, Everett Papers, Series I-5, American Institute of Physics, College Park, MD.
[124] John Wheeler to Allen Shenstone, 28 May 28 1956, WP. *Idem.*

> [Everett's] work suffers from the fundamental misunderstanding which affects all attempts at "axiomatizing" any part of physics. The "axiomatizers" do not realize that every physical theory must necessarily make use of concepts which *cannot*, in principle, be further analysed, since they describe the relationship between the physical system which is the object of study and the means of observation by which we study it: these concepts are those by which we give information about the experimental arrangement, enabling anyone (in principle) to repeat the experiment. It is clear that *in the last resort* we must here appeal to *common experience* as a basis for common understanding. To try (as Everett does) to include the experimental arrangement into the theoretical formalism is perfectly hopeless, since this can only shift, but never remove, this essential use of unanalysed concepts which alone makes the theory intelligible and communicable.

Wheeler came back from Copenhagen defeated, but he had not surrendered. He thought that the dissertation should be approved, but that the battle to convince Bohr should continue, this time by Everett staying in Copenhagen. In fact, the dissertation received formal approval at Princeton in 1957; and an abridged version of it was published in a special issue of *Review of Modern Physics*, along with the proceedings of a conference which Everett never attended, and a note by Wheeler about the possible convergence between Everett's ideas and complementarity. Everett went to Copenhagen in 1959, but the discussions with Bohr bore no fruit. Disillusioned with the whole affair and satisfied with his work on game theory and computers with the Pentagon, Everett abandoned physics, and never again wrote a word on the interpretation of quantum mechanics, even when his ideas were revived ten years later and he had succeeded in a research career related to American defense. Through its first decade of existence, his paper received no more than 20 citations.[126]

---

[125] John Wheeler to Alexander Stern, 25 May 1956. WP, Series 5 – Relativity Notebook 4, p. 92. The sentence is handwritten on the typed letter. It is also written "CWM", which suggests Charles W. Misner was its author. Léon Rosenfeld to Saul M. Bergmann, 21 Dec 1959, RP.

[126] Hugh Everett, (ref. 122); John Wheeler, "Assessment of Everett's 'relative state' formulation of quantum theory," *Rev. Mod. Phys., 29* (1957), 463-465; Hugh Everett, interviewed by C. W. Misner, May 1977. Tape

Blocked by the monocracy around Copenhagen's views on quantum mechanics, Everett's ideas had a fate similar to the causal interpretation: they were poorly received at first, and revived in a different context, many years later. Everett faced some of the obstacles that Bohm had faced earlier, such as the stronghold of the monocracy of the Copenhagen interpretation; the way physics was practiced, with little interest in foundational issues; and the absence of results able to empirically differentiate the alternative interpretations from the complementarity interpretation. He did not face the vicissitudes which Bohm had lived through. One can say that Hugh Everett's case was the opposite of David Bohm's, for contrary to Bohm's exile and persecution by McCarthyism, Everett was working for the Pentagon at the same time that his 'relative states' interpretation of quantum mechanics was suffering an unfavorable reception.

The comparison between Bohm's and Everett's cases suggests that the obstacles they both faced, related to the cultural context of physics, were more influential in the poor reception of their interpretations than were the hardships Bohm faced because the McCarthyist climate. Could the results have been different if the circumstances were different? As fascinating as such a question could be, it is beyond historical investigation. We have seen that the struggles between the supporters of David Bohm and Niels Bohr, struggles in a field clearly dominated by the latter, contributed to creating the frame of a philosophical controversy between the partisans of determinism – causal interpretation - and partisans of the Copenhagen school. The results that Bohm and his colleagues were able to produce – a different epistemology and concepts, and empirical equivalence with non relativistic quantum mechanics, but without empirical or operational differentiation – both produced the disputes of the 1950s and resulted from them. The possibility of producing other results by having more people working on the foundations of quantum mechanics was prevented by the dominance of the Copenhagen school and by the way physics was practiced, i.e. with a low status for foundations of quantum mechanics. Thus, the cultural context emerges as more influential in the poor reception of the causal interpretation than the vicissitudes related the era of McCarthyism and Bohm's exile in Brazil.

THE RUSE OF HISTORY

In the course of time, there has been a change in the context we have been discussing, and Bohm's contribution to this change cannot be underestimated. As I have argued elsewhere, the new context was produced by some new characters, like John Bell, Abner Shimony, John Clauser, Eugene Wigner, Bernard d'Espagnat, Bryce DeWitt, Alain Aspect, and some old ones, like David Bohm and Léon Rosenfeld, but required changes in the physicists' attitude concerning the intellectual and professional status of issues related to the foundations of quantum physics. The monocracy of the Copenhagen school was broken from inside, via a dispute between Wigner and Rosenfeld about the measurement problem in quantum mechanics, in the 1960s. After that point, the literature on the subject has demonstrated a distinction which did not exist in the 1950s, the difference between the Copenhagen school associated with Bohr, and the Princeton school, roughly associated with von Neumann and Wigner. The field of research on foundations of quantum mechanics flourished in the 1970s and the 1980s, especially dealing with issues related to the Bell theorem and the measurement problem. It is a Hegelian ruse of history that the main scientific contribution from this field of research - the Bell theorem and experimental tests confirming quantum mechanical predictions and refuting locality – was motivated by Bohm's insistence on the hidden variables. Honoring Bohm, Bell wrote, "in 1952 I saw the impossible done," referring to Bohm's hidden variable interpretation. I conclude this paper by showing that this statement hides more truth than is usually recognized.[127]

"Smitten by Bohm's paper," the Irish physicist John S. Bell attempted to determine what was wrong with von Neumann's proof, since it did not allow for hidden variables in quantum mechanics. Bell knew von Neumann's proof only indirectly, from his reading of Max Born's *Natural Philosophy of Cause and Chance*, but he could not read von

---

[127] Freire (ref. 3); Olival Freire Jr., "Orthodoxy and heterodoxy in the research on the foundations of quantum physics: E. P. Wigner's case," Boaventura S. Santos, ed., *Cognitive justice in a global world: Prudent knowledge for a decent life*, forthcoming.

Neumann's book because at that time there was no English edition of it. The solution was to ask Franz Mandl, his colleague at Harwell, about the content of the book. "Frank was of German origin, so he told me something of what von Neumann was saying. I already felt that I saw what von Neumann's unreasonable axiom was." He wrote to Pauli asking for reprints of his papers on Bohm's proposal, but he probably did not like the views expressed by Pauli in the de Broglie Festschrift's paper. Bell went to Birmingham in 1953, considering hidden variables as one possibility for his studies. Asked by Rudolf Peierls, who would become his adviser, to give a talk about what he was working on, "Bell gave Peierls a choice of two topics: the foundations of quantum theory or accelerators." Peierls chose the latter, which was the end of the first stage of Bell's involvement with hidden variables. The intermezzo lasted ten years; he only resumed this work at Stanford, during a leave of absence from CERN. In the first of the two articles on the foundations of quantum mechanics that he published while in the U.S., the acknowledgements record both the very origin of his investigation and early and later influences: "The first ideas of this paper were conceived in 1952. I warmly thank Dr. F. Mandl for intensive discussion at that time. I am indebted to many others since then, and latterly, and very especially, to Professor J. M. Jauch." Let us see now what motivation led Bell to resume such issues and the role of Jauch in Bell's work. According to his testimony to Bernstein,

> I had once again begun considering the foundations of quantum mechanics, stimulated by some discussions with one of my colleagues, Josef Jauch. He, it turned out, was actually trying to strengthen von Neumann's infamous theorem. For me, that was like a red light to a bull. So I wanted to show that Jauch was wrong. We had gotten into some quite intense discussions. I thought I had located the unreasonable assumption in Jauch's work.[128]

A few words from the paper published by Jauch and Piron, the motive of discussion with Bell, can enlighten its motivation and complete the picture,

> There are several reasons why we propose to re-examine here von Neumann's proof again. First of all there seems to be a renewed interest in a critique of the foundations of quantum mechanics and some of the recent attempts in this direction have not always done full justice to von Neumann. […] Bohm in his book [*Causality and Chance in Modern Physics*, 1957] even goes so far as to accuse von

---

[128] Jeremy Bernstein, *Quantum Profiles* (Princeton, 1991), 65-68. Wolfgang Pauli to John Bell, 23 Jan 1953, Pauli, (ref. 88), on 28. John Bell, "On the Problem of Hidden Variables in Quantum Mechanics," *Reviews of Modern Physics, 38* (1966), 447-52.

Neumann of circular reasoning. If this were true, this 'proof' would mean, of course, exactly nothing and would leave all doors open for speculations on a 'sub-quantum mechanical level' and a 'deeper reality' so dear to the above-mentioned authors.[129]

In this context, it is very understandable that Bell had introduced his first paper writing that "it is addressed to those who [Jauch] believe that the 'the question concerning the existence of such hidden variables received an early and rather decisive answer in the form of von Neumann's proof on the mathematical impossibility of such variables in quantum theory'." Bell's work can therefore be placed in the tradition related to reinforcing proofs against hidden variables, a tradition that had been challenged by Bohm, de Broglie, and their collaborators. If the possibility of introducing hidden variables in quantum mechanics was the motivation, Bell's approach was far from Bohm's. Indeed, he was not interested in building viable models mimicking quantum mechanics; instead, his works focused on the critical analysis of the assumptions behind von Neumann's proofs and their reformulations, and later on the assumptions behind the Einstein-Podolsky-Rosen *gedankexperiment*. This paper is not the place for presenting the Bell theorem. It is enough for us to note that it contrasted quantum mechanical predictions with a whole family of hidden variables that fulfilled the criterion of locality. To illustrate this criterion, which was implicit in Einstein reasoning, consider a two-particle system. Locality requires that what is being measured on one of the particles does not affect the other. In short, the Bell theorem is the following: no local hidden variable theory can recover all quantum mechanical predictions, and the quantitative measurements of this conflict are the Bell inequalities. This theorem has motivated since the 1970s a cornucopia of experiments aimed at verifying it. Before arriving at this theorem, Bell had shown both the restrictive assumption in von Neumann's proof - the additivity of the expectation values - and why Bohm's hidden variables were possible; they were as nonlocal as quantum theory.[130]

---

[129] J. M. Jauch and C. Piron, "Can Hidden Variables be Excluded in Quantum Mechanics?" *Helvetia Physica Acta 36* (1963), 827-37, on 827. Emphasis in the original

[130] Bell (ref. 128); ibid., "On the Einstein Podolsky Rosen Paradox," *Physics, 1* (1964), 195-200. Michael Stöltzner, "Bell, Bohm, and von Neumann: Some philosophical inequalities concerning no-go theorems and the axiomatic method," Tomasz Placek and Jeremy Butterfield, eds. *Non-locality and modality* (Dordrecht, 2002), 37-58.

Independent of its intrinsic merits, which still awake passions, hidden variables can afford to David Bohm a role in the history of physics comparable to Kepler, who contributed to the creation of modern science while looking for celestial music in the planetary system. In a rough analogy, Newton depended on Kepler as Bell depended on Bohm. The comparison is not mine. In 1958, Lancelot Whyte, an engineer and philosopher of science, defending Bohm against Rosenfeld's attacks, wrote to Rosenfeld, "Naturally you are fully aware […] that valuable results may spring from mistaken motives and reasoning. Kepler is a good example. But this awareness is not evident in your review." Bohm would have enjoyed this comparison, if he had known of it.[131]

---

[131] Lancelot Whyte to Léon Rosenfeld, 8 Apr 1958. RP.

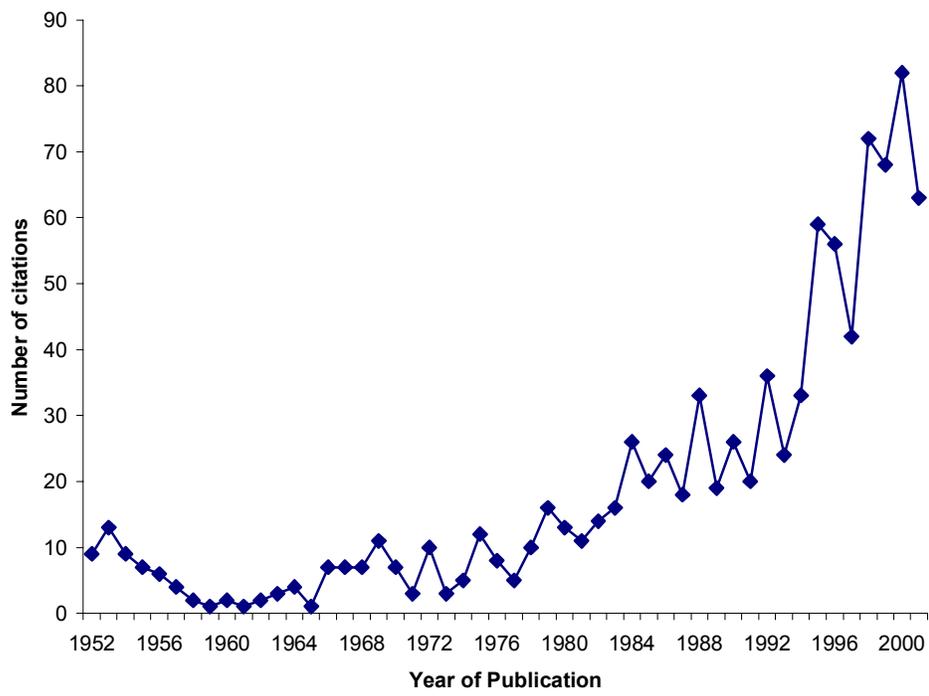

Citations of Bohm, "A suggested interpretation" (*PR*, 1952)

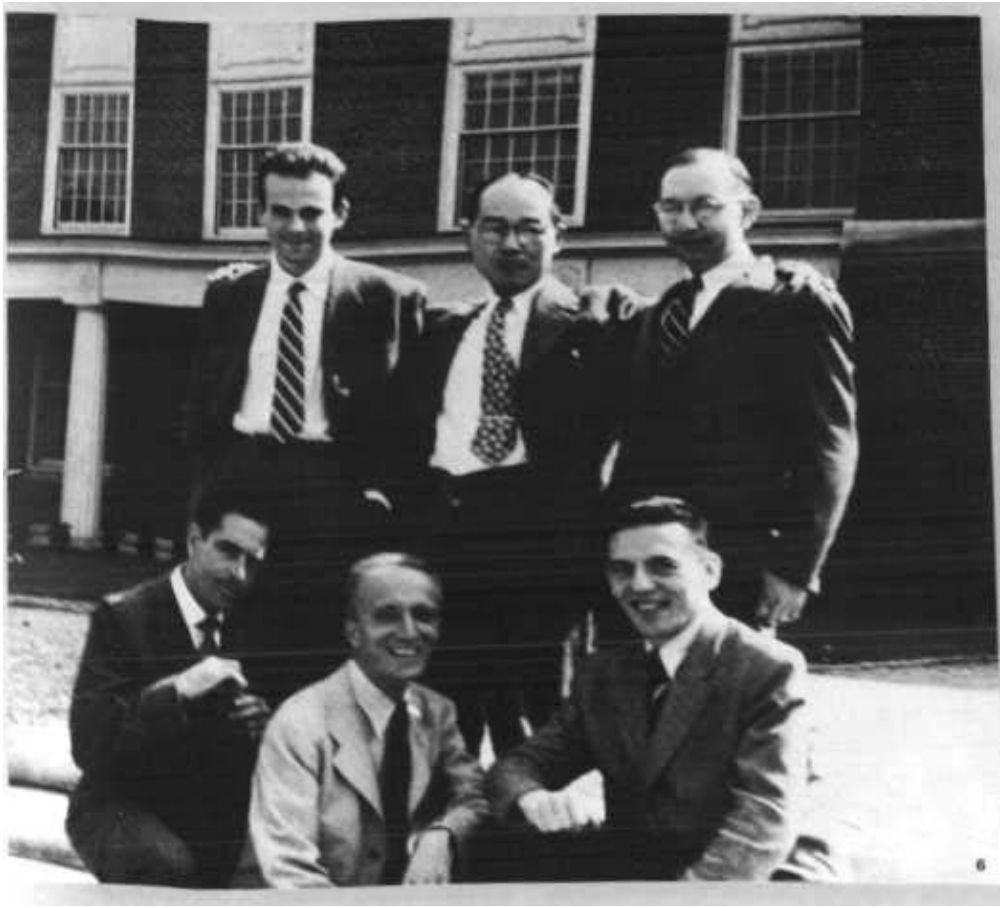

**Cesare Lattes, Hideki Yukawa & Walter Schützer**

**Hervásio Carvalho, José Leite Lopes & Jayme Tiomno**

**Princeton, May, 1949.**

**Source: Leite Lopes Festschrift**